\documentstyle[preprint,floats,prd,aps,epsf]{revtex}
\catcode`@=12
\newcommand{\be}{\begin{equation}}
\newcommand{\ee}{\end{equation}}
\newcommand{\ber}{\begin{eqnarray}}
\newcommand{\eer}{\end{eqnarray}}
\begin{document}
\vspace{0.5in}
\oddsidemargin -.375in  
\newcount\sectionnumber 
\sectionnumber=0 

\def\be{\begin{equation}} 
\def\ee{\end{equation}}
\thispagestyle{empty}
%
%
\vspace {.5in}
\begin{center}
{\Large \bf{CP violation in quasi-inclusive  $B\to K^{(*)} X$ Decays}\\}
\vspace{.5in}
{\rm {T. E. Browder$^1$, A. Datta$^2$, X.-G. He$^3$ and S. Pakvasa$^1$} \\}

\vskip .3in

{\it $^1$Department of Physics and Astronomy, \\}
{\it  University of Hawaii, \\}
{\it Honolulu, Hawaii 96822\\}
{\it $^2$Department of Physics and Astronomy, \\}
{\it  Iowa State University, \\}
{\it Ames, Iowa 50011\\}
{\it $^3$ School of Physics, \\}
{\it  University of Melbourne, \\}
{\it Parkville, Victoria, 3052 \\}
{\it Australia\\}

\vskip .5in
\end{center}  

\begin{abstract}
We consider the possibility of observing CP violation in quasi-inclusive
decays of the type $B^-\rightarrow K^- X$, $B^-\rightarrow K^{*-} X$,
 $\bar B^0\rightarrow K^- X$ and 
$\bar B^0\rightarrow K^{*-} X$, where $X$ does not contain
strange quarks. We present estimates of rates
and asymmetries for these decays in the Standard
Model and comment on the experimental feasibility of observing CP
violation in these decays at future $B$ factories. We find the rate
asymmetries can be quite sizeable. Observation of such asymmetries
could be used to rule out the superweak model of CP violation.
\end {abstract}


\newpage
\baselineskip 24pt

%
\tighten

\section{Introduction}
The possibility of observing large CP violating asymmetries in the
decay of $B$ mesons motivates the construction of high luminosity $B$
factories at several of the world's high energy physics laboratories.
The theoretical and the experimental signatures of these asymmetries
have been extensively 
discussed elsewhere\cite{BABAR},\cite{BELLE},\cite{Br},
\cite{Buras},\cite{dho}. 
At asymmetric $B$ factories, it is possible to measure the
time dependence of 
$B$ decays and therefore time dependent rate asymmetries of
neutral $B$ decays due to $B-\bar B$ mixing. The measurement of time 
dependent asymmetries in the exclusive modes $\bar{B}^0\to \psi K_s$
and $\bar{B}^0\to \pi^+\pi^-$ will allow the determination of the
angles in the Cabbibo-Kobayashi-Maskawa (CKM) unitarity triangle. 
This type of CP violation
has been studied extensively in the literature.

Another type of CP violation also
exists in $B$ decays, direct CP violation in the 
$B$ decay amplitudes. This type of CP violation in $B$ decays has 
also been discussed by several authors although not as extensively.
For charged $B$ decays 
calculation of the magnitudes of the effects for some
exclusive modes and 
inclusive modes have been carried out\cite{Soni},\cite{HouW},
\cite{Wolfie},\cite{FSHe},
\cite{Du},\cite{Kamal},\cite{Kramer}. In contrast with
asymmetries induced by $B-{\overline B}$ mixing, the magnitudes have
large hadronic uncertainties, especially for the exclusive modes.
Observation of these asymmetries can be used to rule out the superweak
class of models\cite{superweak}.

In this paper we describe several quasi-inclusive experimental
signatures which could provide useful information on direct CP violation
at the high luminosity facilities of the future. One of the goals
is to increase the number of events available at experiments for
observing a CP asymmetry. In particular we examine the inclusive
decay of the neutral and the charged $B$ to either a
charged $K$ or a charged $K^*$ meson. By applying the
appropriate cut on the kaon (or $K^*$) energy one can isolate a signal 
with little background from $b\rightarrow c$ transitions.
Furthermore, these quasi-inclusive modes are expected to have less
hadronic uncertainty than
the exclusive modes, would have larger branching ratios and,
compared to the purely inclusive modes they may have larger CP
asymmetries. In this paper 
we will consider modes of the type $B \rightarrow
 K(K^*) X$ that have the strange quark only in the $K(K^*)$-meson.
These processes include contributions from the one loop process
$b\to s g^*\to s q \bar{q}$ as well as the tree 
level process $b\to u  \bar{u} s$. The interference between these
two processes is responsible for the direct CP violation. 

In the next section, we describe
the experimental signature and method. We then calculate
the rates and asymmetries for inclusive $B^-\to K^-(K^{*-})$ and 
$\bar{B}^0\to K^-(K^{*-})$
decays. In the last section, the theoretical uncertainties in the
calculation are discussed.

\section{Experimental Signatures for Quasi-Inclusive $b\to s g^*$}

In the $\Upsilon(4S)$ center of mass frame, 
the momentum of the $K^{(*)-}$ from quasi-two body $B$ decays
such as $B\to K^{(*)-} X$ may have momenta 
above the kinematic limit for $K^{(*)-}$ mesons from 
$b\to c$ transitions.
This provides an experimental signature for $b\to s g^*$, $g^*\to u \bar{u}$ 
or $g^*\to d \bar{d}$ decays where $g^*$ denotes a gluon. 
This kinematic separation between $b\to c$ 
and $b\to s g^*$ transitions is illustrated by
a generator level Monte Carlo simulation in Figure 1 for the case
of $B\to K^{*-}$. (The $B\to K^-$ spectrum will be similiar).
This experimental signature can be applied to the asymmetric energy
$B$ factories if one boosts backwards along the z axis into the 
$\Upsilon(4S)$ center of mass frame.

Since there is a large background (``continuum'')
from the non-resonant processes $e^+ e^-\to q \bar{q}$ where
$q=u, d, s, c$, experimental cuts on the event shape are also imposed. 
To provide additional continuum suppression, the  
``$B$ reconstruction'' technique has been employed. The requirement
that the kaon and $n$ other pions form a system consistent in beam
constrained mass and energy with a 
$B$ meson dramatically reduces the background. After these
requirements are imposed, one searches for an excess in the kaon
momentum spectrum above the $b\to c$ region. 
Only one combination per event is chosen.
No effort is made to unfold the feed-across
between submodes with different values of $n$.

\begin{figure}[htb]
\centerline{\epsfysize 3.4 truein \epsfbox{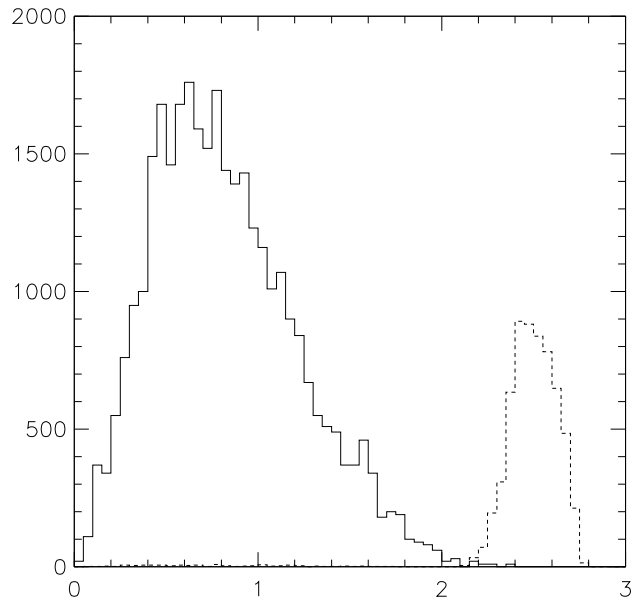}}
\caption{Generated inclusive $B\to K^{*-}$ momentum spectrum.
The component below $2.0$ GeV/c is due to $b\to c$ decays while the
component above 2.0 GeV/c arises from quasi-two body $b\to s g^*$ decay.
The normalization of the $b\to c$ component is reduced by a factor of
approximately 100 so that both components are visible.}
\label{bkx_inclusive}
\end{figure}

Methods similar to these have been successfully used 
by the CLEO~II experiment to isolate a signal
in the inclusive single photon energy spectrum
and measure the branching fraction for inclusive $b\to s\gamma$ transitions
and to set upper 
limits on $b\to s \phi$ transitions\cite{cleobsg},\cite{cleophix}. 
It is clear from these studies that the $B$ reconstruction method provides
adequate continuum background suppression.

The decay modes that will be used here are listed below:
\begin{enumerate}
\item $B^-\to K^{(*)-}\pi^0$
\item $\bar{B}^0\to K^{(*)-} \pi^+$ 
\item $B^- \to K^{(*)-} \pi^-\pi^+$ 
\item $\bar{B}^0\to K^{(*)-}\pi^+\pi^0$
\item $\bar{B}^0\to K^{(*)-}\pi^+\pi^-\pi^+$
\item $B^-\to K^{(*)-} \pi^+\pi^-\pi^0$
\item $B^-\to K^{(*)-}\pi^+\pi^-\pi^+\pi^-$
\item $\bar{B}^0\to K^{(*)-}\pi^+\pi^-\pi^+\pi^0$
\end{enumerate}

In case of multiple entries for a decay mode, we choose the
best entry on the basis of a $\chi^2$ formed from the beam
constrained mass and energy difference
[i.e. $\chi^2= (M_B/\delta M_B)^2+ (\Delta E/\delta\Delta E)^2$]. 
In case of multiple
decay modes per event, the best decay mode candidate is picked
on the basis of the same $\chi^2$.

\begin{figure}[htb]
\centerline{\epsfysize 5.0 truein \epsfbox{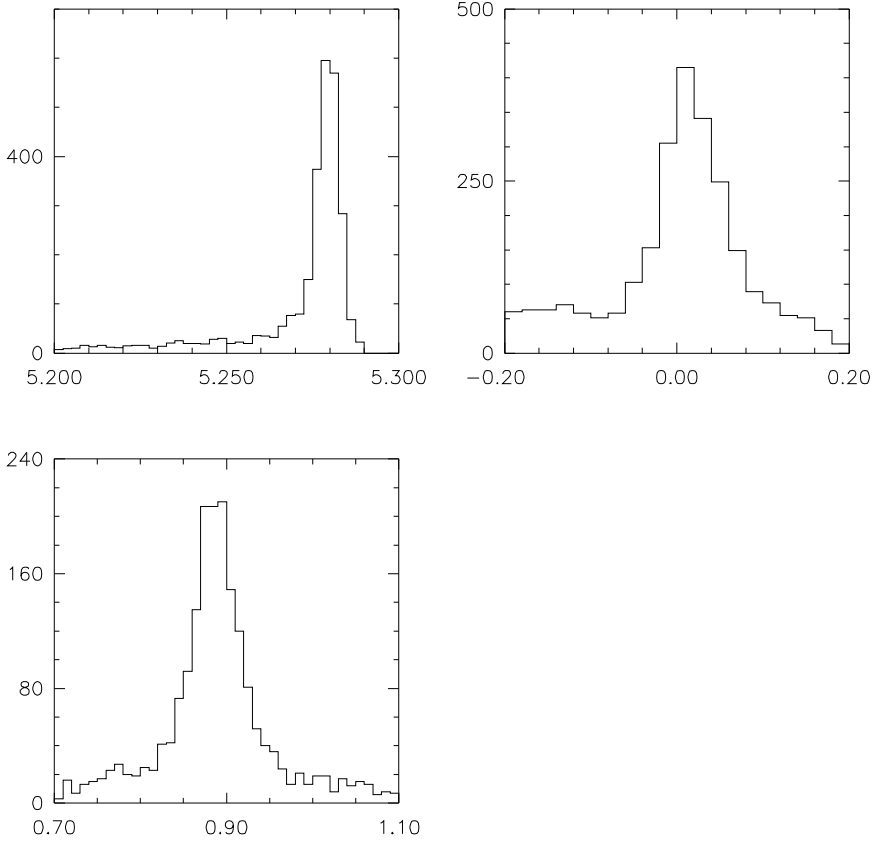}}
\caption{Monte Carlo simulation of the inclusive $B\to K^{*-} X$ signal
with the $B$ reconstruction method: (a) the beam constrained mass
distribution (b) the distribution of energy difference (c) the $K^-
\pi^0$ invariant mass after selecting on energy difference and beam
constrained mass}
\label{bkx_brecon}
\end{figure}

Cross-feed between different $b\to s g$ decay modes 
(i.e. the misclassification of decay modes)  provided
the $K^{(*)-}$ is correctly identified, is not a concern 
as the goal is to extract an inclusive signal
(an example of the signal is shown in Fig.~\ref{bkx_brecon}).
The purpose of the
$B$ reconstruction method is to reduce continuum background.
As the multiplicity of the decay mode increases, however,
the probability of misrecontruction will increase.

The signal is isolated as excess $K^{(*)-}$ production in the high
momentum signal region ($2.0<p_{K^{(*)}}<2.7$ GeV) above 
the continuum background.
To reduce contamination from high momentum 
$B\to \pi^- (\rho^-)$ production and
residual $b\to c$ background, we assume the presence of a high
momentum particle identification system as will be employed in
the BABAR, BELLE, and CLEO~III experiments.

We propose to measure the asymmetry
$N(K^{(*)+} - K^{(*)-})/N(K^{(*)+} + K^{(*)-})$ where $K^{(*)\pm}$ originates
from a partially reconstructed $B$ decay such as $B\to K^{(*)-}
(n\pi)^0$ with $2.7>p(K^{(*)-})>2.0$ GeV. 
The additional pions have net charge $0$, $n\le 4$ and at most one neutral
pion is allowed. 
We assume that the contribution from $B\to K^-\eta^{'} X $ 
decays has been removed by cutting on the $\eta^{'}$ region in
$X$ mass. It is possible that the anomalously large rate from
this source\cite{cleo_etaprime} could dilute the asymmetry.

\section{ Effective Hamiltonian} 
In the Standard Model (SM) 
the amplitudes for hadronic $B$ decays of the type $b\to q \bar{f} f$ 
are generated by the following effective 
Hamiltonian \cite{Reina}:
\begin{eqnarray}
H_{eff}^q &=& {G_F \over \protect \sqrt{2}} 
   [V_{fb}V^*_{fq}(c_1O_{1f}^q + c_2 O_{2f}^q) -
     \sum_{i=3}^{10}(V_{ub}V^*_{uq} c_i^u
+V_{cb}V^*_{cq} c_i^c +V_{tb}V^*_{tq} c_i^t) O_i^q] +H.C.\;,
\end{eqnarray}
where the
superscript $u,\;c,\;t$ indicates the internal quark, $f$ can be $u$ or 
$c$ quark. $q$ can be either a $d$ or a $s$ quark depending on 
whether the decay is a $\Delta S = 0$
or $\Delta S = -1$ process.
The operators $O_i^q$ are defined as
\begin{eqnarray}
O_{f1}^q &=& \bar q_\alpha \gamma_\mu Lf_\beta\bar
f_\beta\gamma^\mu Lb_\alpha\;,\;\;\;\;\;\;O_{2f}^q =\bar q
\gamma_\mu L f\bar
f\gamma^\mu L b\;,\nonumber\\
O_{3,5}^q &=&\bar q \gamma_\mu L b
\bar q' \gamma^\mu L(R) q'\;,\;\;\;\;\;\;\;O_{4,6}^q = \bar q_\alpha
\gamma_\mu Lb_\beta
\bar q'_\beta \gamma^\mu L(R) q'_\alpha\;,\\
O_{7,9}^q &=& {3\over 2}\bar q \gamma_\mu L b  e_{q'}\bar q'
\gamma^\mu R(L)q'\;,\;O_{8,10}^q = {3\over 2}\bar q_\alpha
\gamma_\mu L b_\beta
e_{q'}\bar q'_\beta \gamma^\mu R(L) q'_\alpha\;,\nonumber
\end{eqnarray}
where $R(L) = 1 \pm \gamma_5$, 
and $q'$ is summed over u, d, s, c and b.  $O_2$ and $O_1$ are the tree
level and QCD corrected operators, respectively. 
$O_{3-6}$ are the strong gluon induced
penguin operators, and operators 
$O_{7-10}$ are due to $\gamma$ and Z exchange (electroweak penguins),
and ``box'' diagrams at loop level. The Wilson coefficients
 $c_i^f$ are defined at the scale $\mu \approx m_b$ 
and have been evaluated to next-to-leading order in QCD.
The $c^t_i$ are the regularization scheme 
independent values obtained in Ref. \cite{FSHe}.
We give the non-zero  $c_i^f$ 
below for $m_t = 176$ GeV, $\alpha_s(m_Z) = 0.117$,
and $\mu = m_b = 5$ GeV,
\begin{eqnarray}
c_1 &=& -0.307\;,\;\; c_2 = 1.147\;,\;\;
c^t_3 =0.017\;,\;\; c^t_4 =-0.037\;,\;\;
c^t_5 =0.010\;,
 c^t_6 =-0.045\;,\nonumber\\
c^t_7 &=&-1.24\times 10^{-5}\;,\;\; c_8^t = 3.77\times 10^{-4}\;,\;\;
c_9^t =-0.010\;,\;\; c_{10}^t =2.06\times 10^{-3}\;, \nonumber\\
c_{3,5}^{u,c} &=& -c_{4,6}^{u,c}/N_c = P^{u,c}_s/N_c\;,\;\;
c_{7,9}^{u,c} = P^{u,c}_e\;,\;\; c_{8,10}^{u,c} = 0
\end{eqnarray}
where $N_c$ is the number of color. 
The leading contributions to $P^i_{s,e}$ are given by:
 $P^i_s = ({\frac{\alpha_s}{8\pi}}) c_2 ({\frac{10}{9}} +G(m_i,\mu,q^2))$ and
$P^i_e = ({\frac{\alpha_{em}}{9\pi}})
(N_c c_1+ c_2) ({\frac{10}{9}} + G(m_i,\mu,q^2))$.  
The function
$G(m,\mu,q^2)$ is given by
\begin{eqnarray}
G(m,\mu,q^2) = 4\int^1_0 x(1-x)  \mbox{ln}{m^2-x(1-x)q^2\over
\mu^2} ~\mbox{d}x \;.
\end{eqnarray}
All the above coefficients are obtained up to one loop order in electroweak 
interactions. The momentum $q$ is the momentum carried by the virtual gluon in
the penguin diagram.
When $q^2 \rangle  4m^2$, $G(m,\mu,q^2)$ becomes imaginary. 
In our calculation, we 
use $m_u = 5$ MeV, $m_d = 7$ MeV, $m_s = 200$ MeV, $m_c = 1.35$ GeV
\cite{lg,PRD}.

We assume that the final state phases calculated at the quark level
will be a good approximation
 to the sizes and the signs of the final state interaction
(FSI) phases at
the hadronic level for quasi-inclusive decays
when the final state particles are quite energetic
as is the case for the $B$ decays in the kinematic range
of experimental interest\cite {Soni}.
As pointed out by Gerard and Hou\cite{HouW} and 
clarified by Wolfenstein\cite{Wolfie}. 
when calculating rate asymmetries
using the absorptive amplitude given above,  
one must be careful 
to be consistent with the requirements of the CPT theorem. 
Gerard and Hou\cite{HouW} noted that CPT is
violated if one does not include all diagrams of the same
order. The interference term responsible for the rate asymmetry
due to ${c_i}^u$ contains two contributions:
an interference between penguin amplitudes of order 
${\alpha_s}^2$ and a contribution
from the interference of the tree amplitude with a higher order penguin
diagram that contains an absorptive part from 
a vacuum polarization bubble in the gluon propagator. These two 
contributions cancel each other.
Therefore, in practical calculations ${c_i}^u$ must be
treated as real. The general rule is that the phase of the penguin Wilson
coefficient 
must be dropped if there is a tree amplitude with the same CKM factor
and the final states for the tree and penguin amplitudes are the
same. In a more general analysis of this problem from CPT and
unitarity consideration\cite{Wolfie}, 
Wolfenstein  showed that diagonal strong
phases (the phases due to the rescattering of the state which is the
same as the final state e.g. $u \bar{u}\to u \bar{u}$) 
do not contribute to partial rate asymmetries. 
The phase in ${c_i}^u$ is a diagonal phase in this sense.
We will follow this prescription of Ref.~\cite{HouW} 
to remove the redundant strong phases.

\section{Matrix Elements for $ B^-  \rightarrow {K^-}  X $
and $\bar{B}^0  \rightarrow {K^-}  X $}

We proceed to calculate 
 the matrix elements of the form $\langle K X|H_{eff}|B\rangle $ which represents 
the process  $ B\rightarrow K X$ and where $H_{eff}$ has been described 
above. The effective Hamiltonian consists of operators with a
 current $\times$ current structure. 
Pairs of such operators can be expressed in terms of
color singlet and color octet structures which lead to color singlet and
color octet matrix elements. In the factorization approximation,
one separates out the currents in the operators by inserting the vacuum
state and neglecting any QCD interactions between the two currents. The
basis for this approximation is that, if the quark pair created by one of
the currents carries large energy then it will not have significant 
QCD interactions. In this approximation the color octet matrix element
does not contribute because it cannot be expressed in a factorizable
color singlet form. The color octet operators could contribute if, for
instance, the
 quark pair emits or absorbs a gluon\cite{oct}. It has been shown\cite{Don}
that in the leading order, where the energy of the light quark pair
$E\sim m_b$ with $m_b \rightarrow \infty$, the octet matrix element
vanishes when the final state is two $0^-$ mesons.
In our case, since the
energy of the quark pairs that either creates the $K$ or the $X$ state
is rather large, factorization is likely to be a good first 
approximation. To accommodate some 
deviation from this approximation we treat
$N_c$, the number of colors that enter in the calculation of the matrix
elements, as a free parameter. In our calculation we will see how our
results vary with different choices of $N_c$. The value of $N_c\sim 2$ is
suggested by experimental data on 
low multiplicity hadronic $B$ decays\cite{Br}.

In
the factorization approximation 
the matrix element of $B^- \rightarrow K^-X$ decay can be expressed as, 
\be
	M = M_1 + M_2
\ee
where
\ber
	M_1 & = &\frac{G_F}{\protect \sqrt{2}}
             \langle K^-|\, \bar{s} \gamma^\mu(1-\gamma^5) \, b\, |\, B \rangle
	      \sum_{q=u,d,s} 
              \langle X\, |\, \bar{q} \, \gamma_\mu \{L_q (1-\gamma^5)
	      + R_q (1+\gamma^5) \} \, q\, |\, 0\rangle \nonumber  \\
	M_2 & = &\frac{G_F}{\protect \sqrt{2}} (FL_{u} 
             \langle X\,|\,\bar{u} \,\gamma^\mu (1-\gamma^5) \,b\,|\,B\rangle 
	      \langle K^-|\,  \bar{s} \, \gamma^\mu (1-\gamma^5) \,
               u\, |\, 0\rangle  
		\nonumber \\
	    &   &  + FR_{u}
          \langle X\, |\, \bar{u} \, (1-\gamma^5) \, b\, |\, B\rangle  
	      \langle K^-|\, \bar{s} \, (1+\gamma^5) \, u\, |\, 0\rangle  )
\eer
where
\ber
	L_u & = & V_u \left(c_1 + \frac{c_2}{N_c} \right) + A_3 
		+ \frac{1}{N_c} A_4 + A_9 + \frac{1}{N_c} A_{10}\nonumber  \\
	L_d & = & A_3 + \frac{1}{N_c} A_4 - \frac{1}{2} 
		\left(A_9 + \frac{1}{N_c} A_{10} \right)\nonumber \\ 
	L_s & = & A_3 + \frac{1}{N_c} A_4 - \frac{1}{2} 
		\left(A_9 + \frac{1}{N_c} A_{10} \right)\nonumber \\
	FL_{u} & = & V_u \left(\frac{c_1}{N_c} + c_2 \right) + 
                  \frac{A_3}{N_c} + A_4
		+ \frac{A_9}{N_c} + A_{10} \nonumber\\
	R_u & = & A_5 + \frac{1}{N_c} A_6 + A_7 + \frac{1}{N_c} 
A_8 \nonumber \\
	R_d & = & A_5 + \frac{1}{N_c} A_6 
		- \frac{1}{2} \left(A_7 + \frac{1}{N_c} A_8 \right)\nonumber \\
	 R_s  & = & A_5 + \frac{1}{N_c} A_6 
		- \frac{1}{2} \left(A_7 + \frac{1}{N_c} A_8 \right)\nonumber \\
	FR_{u} & = & -2 \, \left(\frac{1}{N_c} A_5 + A_6 + \frac{1}{N_c} A_7 
		+ A_8 \right)
\eer
We have defined
\be
	A_i = - \sum_{q=u,c,t} c_i^q V_q
\ee
with
\be
	V_q = V_{qs}^{*} V_{qb}
\ee

Using the definition
\be
	\langle K^-|\, \bar{s} \, \gamma^\mu (1-\gamma^5) \, u\, |\, 0\rangle  
	= i f_K p_K^\mu
\ee
where $f_K$ is the kaon decay constant, and the free quark equation 
of motion one has
\be
	\langle K^-|\, \bar{s} \, (1+\gamma^5) \, u\, |\, 0\rangle 
	= -i \frac{f_K m_K^2}{m_u + m_s}
\ee
Using these two results we can simplify $M_2$ and write it in the form
\be
	M_2 = i f_K [\, \alpha 
                 \langle X\, |\, \bar{u} \, (1+\gamma^5) \, b\, |\, B^-\rangle 
 + \, \beta \langle X\, |\, \bar{u} \, (1-\gamma^5) \, b\, |\, B^-\rangle ]
\ee
with
\ber
	\alpha & = & m_b FL_{u}\nonumber \\
	\beta & = & - m_u FL_{u} - \frac{FR_{u} m_K^2}{m_s + m_u}
\eer
To calculate $M_1$ we express
\be
	\langle K^-|\, \bar{s} \,\gamma^\mu (1-\gamma^5) \, b\, |\, B\rangle 
	= {f_+} (p_B + p_K)^\mu + {f_-} (p_B - p_K)^\mu
\ee
where ${f_+}, {f_-}$ are Lorentz invariant form factors which are 
functions of $(p_B - p_K)^2$.

For the decay $\bar{B}^0 \rightarrow K^- X$ decay, $M_1 = 0$ and only
$M_2$ contributes.

\section{Matrix Elements for
 $ B^- \to K^{*-} X $  and $\bar{B}^0 \to K^{*-} X $ }
	
For $ B^- \rightarrow {K^{*-}} X$ decay, we also write the matrix element as
\be
	M = M_1 + M_2
\ee
where $M_1$ has a similar structure as in Eq. (6). $M_2$ has the form
\be
	M_2 = FL_{u} \langle X|\bar{u} 
                \gamma^\mu (1-\gamma^5) b | \bar{B} \rangle 
	      m_{K^*} g_{K^*} \varepsilon_\mu^\lambda
\ee
where
\be
	\langle {K^\lambda}^* |\, \bar{s} \, \gamma^\mu (1-\gamma^5) \, u\, |\,  0 \rangle 
	= m_{K^*} g_{K^*} {{\varepsilon_\mu}^\lambda}^*
\ee
with $g_{K^*}$, the decay constant and ${{\varepsilon_\mu}^\lambda}^*$ being
the polarization vector of the vector meson.

For the $\bar{B}^0 \rightarrow K^{*-} X$ decay only $M_2$ contributes. 
To calculate
$M_1$, following the notation of Bauer, Stech
and Wirbel (BSW) \cite{BSW} we write
\ber
	\langle K^{*-} |\, \bar{s} \, 
        \gamma^\mu (1-\gamma^5) \, b \, |\,B^- \rangle 
	& = & b_1 \varepsilon_{\mu\alpha\beta\gamma} {\varepsilon^\alpha}^*
	p_B^\beta p_{K^*}^\gamma  \\
	& & + i \{ b_2 {\varepsilon^\mu}^* 
	- b_3 (p_B + p_{K^*})_\mu \varepsilon^* \cdot k 
	+ b_4 \varepsilon^* \cdot k k_{\mu} \}
\eer
with
\ber
	b_1 & = & \frac{2V(k^2)}{m_B + m_{K^*}} \\
	b_2 & = & (m_B + m_{K^*}) A_1(k^2) \\
	b_3 & = & \frac{A_2(k^2)}{m_B + m_{K^*}} \\
	b_4 & = & \frac{2m_{K^*} [A_0(k^2) - A_3(k^2)] }{k^2}
\eer
where $k=p_B-p_{K^*}$.
 In our calculation we will use the form factors of
Ref\cite{Narduli} for the primary result.
To check the dependence of the results on
form factors we will also 
use the modified BSW model\cite{BSW1} which has a
 dipole  behavior for the form factors.
The form factors in
Ref\cite{Narduli} are first evaluated at $k^2=0$ and then extrapolated
  to a finite $k^2$ assuming 
a monopole behavior for all the form factors.

 We considered
the possible contribution from annihilation graphs to both decay
rates and asymmetries.
In agreement with previous estimates, 
the annihilation contribution to rates is found to be 
small \cite{Gron}. The contribution to CP asymmetries is potentially
interesting since the dependence on CKM parameters is quite different.
In this case if we limit ourselves to the processes
$b\to s u \bar{u}$, $b\to  s d \bar{d}$
that have only a strange quark
in the $K^{(*)-}$ meson then the contribution to the
asymmetry from the annihilation term
turns out to be too small to be of interest.


\section{Decay distribution and CP asymmetries}
In this section we describe the formalism to calculate the decay 
distribution, asymmetries and the decay rates.

The general form of the matrix element is 
\be
        M = M_1 + M_2
\ee
and so
\be
	|M|^2 = |M_1|^2 + |M_2|^2 + M_1^{\dagger} M_2 + M_1 M_2^{\dagger}
\ee
Now $|M_1|^2$ has the structure
\be
	|M_1|^2 = H_{\mu\nu} W^{\mu\nu}
\ee
where
\be
	H_{\mu\nu} = \langle K^{-}, ({K^{*-}}) |\,  J_\mu \, | B^{-} \rangle 
		     \langle  B^{-} |\, {J_\nu}^\dagger \, | K^{-}, ({K^{*-}}) \rangle 
\ee
and
\be
	W_{\mu\nu} = \sum_X (2 \pi)^4 \delta^4(p_B - p_K -p_X)
		 \langle 0|J_\mu|X\rangle \langle X| {J_\nu}^\dagger |0\rangle 
\ee
with
\be
	J_\mu = \sum_{u,d,s} \bar{q} \gamma_\mu 
		\{L_q (1-\gamma^5) + R_q (1+\gamma^5)\} q
\ee
In the parton model approximation we can interpret the above process as
the decay
\be \label{eq:star}
	B (p_B) \rightarrow K (p_K) + q (p_1) + \bar{q} (p_2)
\ee
with $p_X = p_1 + p_2$. 

We can also express
\be
	W_{\mu\nu} = 2 Im ~ i \int d^4x e^{-iqx}
		 \langle 0\, | 
T [ J_\mu(x) J_\nu^\dagger(0) ] |\, 0\rangle 
\ee
with $q=p_B-p_K=p_X$. The parton model approximation is the leading term
in the expansion for the $T$ product in the above equation
 and so this form for $W_{\mu\nu}$ is useful to
calculate higher order corrections to the parton model approximation.

The decay distribution is given by
\be
	\frac{d\Gamma}{dE_K} = \frac{1}{(2\pi)^3} \frac{1}{16m_B^2}
	\int |M_1|^2 dm_{12}^2
\ee
where $m_{12}^2 = (p_1 + p_K)^2$ and $|M_1|^2$ has the structure
\be
	|M_1|^2 = 24 \sum_{u,d,s} 
\{ [ (|L_q|^2 + |R_q|^2) A_{\mu\nu}^1 +(|L_q|^2 - |R_q|^2) A_{\mu\nu}^2 ]
 H^{\mu\nu}
	         - 2 m_q^2 g_{\mu\nu} H^{\mu\nu} Re(L_q R_q^*) \}
\ee
with
\ber
	A^1_{\mu\nu} & 
= & p_{1\mu} p_{2\nu} + p_{1\nu} p_{2\mu} 
	   	     - g_{\mu\nu} p_1 \cdot p_2 \nonumber\\
A^2_{\mu\nu} &= & i\epsilon_{\mu \alpha \nu \beta}p^{2 \alpha}p^{1 \beta}\\ 
\eer
(See the Appendix for the full form of $|M_1|^2$.)

For decays involving $K^-$, $|M_2|^2$ has the form
\ber
	|M_2|^2 & = & \sum_X \left| \langle X\, |\,  \bar{u} \, 
              \{ \alpha (1+\gamma^5) 
		+ \beta (1-\gamma^5) \} \, b \, |\, B\rangle  \right|^2
		(2\pi)^4 \delta^4(p_B-p_K-p_X) \\
	& = & \sum_X (2\pi)^4 \delta^4(p_B-p_K-p_X) 
             \langle B\, |\, J\, |\, X\rangle 
		\langle X\, |\, J^\dagger \, |\, B\rangle 
\eer
with
\be
	J^\dagger = \bar{u} \{ \alpha (1+\gamma^5) + \beta (1-\gamma^5) \} b
\ee

In the parton model approximation we replace
\be
	\sum_X |X \rangle 
          \langle X| \rightarrow \sum_s \int\frac{d^3p}{(2\pi)^3
           2E_u}|u(p_u,s)\rangle \langle u(p_u,s)| \\
\ee
where $|u(p_u,s)\rangle $ is a free quark state 
with momentum $p_u$ and spin $s$. As in the previous case
 it is also possible to express $|M_2|^2$ as
\be \label{eq:m22}
	|M_2|^2 = 2 Im 
               \langle B|i\int d^4x e^{ip_Kx} T [ J(x)
J^\dagger(0) ] |B\rangle 
\ee
where the parton model approximation is again the leading term in the
expansion of the $T$ product in the above equation and can be
interpreted as the
two body process
$b\rightarrow K u $.
 
In the parton model approximation we can write for $K^-$ decay,
\be
	|M_2|^2 = 4 f_K^2 \left[(|\alpha|^2+|\beta|^2) p_b \cdot p_u
		+ 2 Re (\alpha \beta^*) m_b m_u \right]
\ee
and for $K^{*-}$ decay
\be
	|M_2|^2 = 4 |F_{Lu}|^2 m_{K^*}^2 g_{K^*}^2 \left[  p_b \cdot p_u
	+ \frac{2 p_b \cdot p_{K^*} p_{K^*}\cdot p_u }
		{M_{K^*}^2} \right]
\ee

For the interference terms, we have, for the $K^-$ decay
\ber
 M_1 M_2^\dagger & = &-i f_K
     \langle K^-|\bar{s}\gamma^\mu(1-\gamma^5)b|B\rangle   \nonumber\\
	& & 
    \sum_X \langle B|\bar{b} \{\alpha^* (1-\gamma^5) + \beta^* (1+\gamma^5) \}
        u|X\rangle \nonumber\\
& &	\langle X|{\bar{u}} \gamma_\mu
        \{ L_u (1-\gamma^5) + R_u (1+\gamma^5) \} u|0
         \rangle (2\pi)^4 \delta^4(p_B-p_K-p_X) 
\eer
In the parton model approximation this is written as [using Eq. (39)]
\ber
 M_1 M_2^\dagger &=& -i f_K
\langle K^-|\bar{s}\gamma^\mu(1-\gamma^5)b|B\rangle   \nonumber\\
	& & \int 
        \langle B|\bar{b} \{\alpha^* (1-\gamma^5) + \beta^* (1+\gamma^5) \}
	( {\not p}_u + m_u)\nonumber\\ 
& &     \gamma_\mu 
	\{ L_u (1-\gamma^5) + 
R_u (1+\gamma^5) \} u|0\rangle  \frac{d^3 p_u}{(2\pi)^3 2E_u}
(2\pi)^4 \delta^4(p_b-p_K-p_u) 
\eer
Kinematically this term looks like the two body decay 
$b\rightarrow K u $.

Using the definition 
\be
	\langle B^-|\bar{b} \gamma^\mu (1-\gamma^5) u|0\rangle  = i f_B p_B^\mu
\ee
and the quark equation of motion, we have
\be
	\langle B^-|\bar{b} (1+\gamma^5) u|0\rangle  = - i f_B \frac{m_B^2}{m_b+m_u}
\ee
and finally we can write (dropping the u quark phase space factor and
the delta function)
\be
	M_1 M_2^\dagger + M_1^\dagger M_2 = 2 f_B f_K 
	[Re C \{ g_+ p_B \cdot p_u + g_- p_K \cdot p_u\} 
	+ Re D \{ g_+ m_B^2 + g_- p_B \cdot p_K\} ]
\ee
\ber
        g_+ & = & f_+ + f_- \\
        g_- & = & f_+ - f_- \\
	C   & = & \frac{2 m_B^2}{m_b+m_u} (\alpha^* L_u - \beta^* R_u) \\
	D   & = & [ \beta^*L_u -\alpha^*R_u)] 2 m_u
\eer
For decay to $K^{*-}$ one can write a similar expression
\ber
	M_1 M_2^\dagger + M_1^\dagger M_2 & = &
		4 Re(FL_u^* L_u)m_{K^*} g_{K^*} f_B\sum_i x_i + \nonumber\\
              &  &4 Re(FL_u^* R_u)m_{K^*} g_{K^*} f_B\frac{m_um_B^2}{m_b+m_u}
\sum_i y_i \ 
\eer
where the expressions for $x_i,y_i$ are given in the
Appendix .

There will be higher order perturbative corrections, such as additional 
gluons in the final states, to the processes described
above. These effects are
expected to be small. We will, however, include the bound state effects
of the b quark inside the $B$ meson on the decay distributions .

Inside the $B$ meson, the $b$ quark is not on-shell. This will cause 
the energy to have a distribution even in the case of "two body" decay. 
To obtain the decay distribution we consider a model for the $B$-meson
wavefunction \cite{Model} which has been
used for the calculation of the photon spectrum of
inclusive $b \rightarrow s \gamma$ decays in Ref.~\cite {ABS}.  

In the rest frame of the B-meson, the b-quark and the light antiquark
$\bar{q}$ inside the $B$ meson  with energies $E_b$ and $E_q$ satisfy
\ber\label{eq:mb}  
	E_b + E_q & = & \sqrt{p^2 + m_q^2} + \sqrt{p^2 + m_b^2} \\
	          & = & m_B 
\eer
where $\vec{p_b} = -\vec{p_{\bar{q}}}, |\vec{p_b}|=p=|\vec{p_{\bar{q}}}|$.

To satisfy Eq. (53) for 
all $p$ the $b$-quark mass is considered a function of $p$,
\be
	m_b^2 = m_B^2 + m_{\bar{q}}^2 - 2 m_B E_q
\ee
The $B$-meson wave function is taken as
\be
	\phi(p) = \frac{4}{\protect\sqrt{\pi}} \frac{1}{p_F^3}
		e^{-\frac{p^2}{p_F^2}}
\ee
with the normalization
\be
	\int_0^\infty p^2 \phi(p) dp = 1.
\ee
For our numerical results we will use $m_{u,d} = 150$ MeV 
and the Fermi momentum $p_F=0.3$ GeV.

The decay distribution is now obtained from
\be
	\frac{d\Gamma}{dE_K} = \int p^2 dp \phi(p)
		\left(\frac{d\Gamma}{dE_K}\right)_{partonic}
\ee
where the partonic distribution 
$(\frac{d\Gamma}{dE_K})_{partonic}$ can be obtained by boosting the decay
distribution in the rest frame of the b-quark to the rest frame of the
B-meson.

To complete the numerical calculations we have to fix the value of
the gluon momentum $q^2$ in the G function of Eq. (4). 
For the ``three body'' decays governed by $|M_1|^2$, 
$q^2=p_X^2$ 
while for the ``two body'' decay governed by $|M_2|^2$ and the
interference terms  one can use simple 
two body kinematics \cite{Desh} to obtain $q^2\approx m_b^2/2$. In our
calculation of $M_2$ we compare results with
 $q^2 = m_b^2/3$ to those with 
$q^2= m_b^2/2$ in order to assess the dependence of the final result
on this uncertainty.

Having obtained the decay distributions we define the asymmetry for
a $B$ decay,
\be
	a = \frac{ \frac{d\Gamma}{dE_k} (B \rightarrow K^{+(*)} X)
		- \frac{d\Gamma}{dE_k} (\bar{B} \rightarrow K^{-(*)} X)}
		{ \frac{d\Gamma}{dE_k} (B \rightarrow K^{+(*)} X)
		+ \frac{d\Gamma}{dE_k} (\bar{B} \rightarrow K^{-(*)} X)}
\ee
Integrated decay
rates and integrated partial rate asymmetries can also be obtained in
the usual manner.

Following Ref~\cite{HouW} we can write the amplitudes, both $M_1$ and
$M_2$ as
\ber
M_{i} &=& V_u{A_{u}^{i}} +V_c{A_{c}^{i}} +V_t{A_{t}^{i}}\nonumber\\
    &=& V_u{\Delta_{ut}^{i}} +  V_c{\Delta_{ct}^{i}}\
\eer
where $i=1, 2$ and 
\ber
{\Delta_{ut}^{i}} & = & {A_{u}^{i}} - {A_{t}^{i}}\nonumber\\ 
{\Delta_{ct}^{i}} & =&  {A_{c}^{i}} - {A_{t}^{i}}\
\eer
and the unitarity relation of the CKM, $V_u +V_c + V_t=0$ has been used.
Note that 
${\Delta_{ut}^{i}}$ also contains the tree level amplitude.

The decay distributions are such that the contribution from $M_1$
and $M_2$ are separated in $E_{K}$ in the approximation of
neglecting the interference term. 
The decay distribution is a sum of two independent decay
distributions governed by $M_1$ and $M_2$. From the
structure of $M_1$ and $M_2$ it can be seen that the Wilson
coefficients that occur in
$M_1$ and $M_2$ contribute in pairs of the type $c_i +c_{i+1}/N_c$
and $c_{i+1} +c_{i}/N_c$, respectively. 
The values of the  Wilson's coefficients are
such that generally the first combination is suppressed relative to the 
second and hence $M_2$ is enhanced relative to $M_1$. Thus the
decay distribution associated with
 $M_2$ is larger than the decay distribution associated with $M_1$.

\begin{figure}[htb]
\vskip 0.9in
\centerline{\epsfysize 3.0 truein \epsfbox{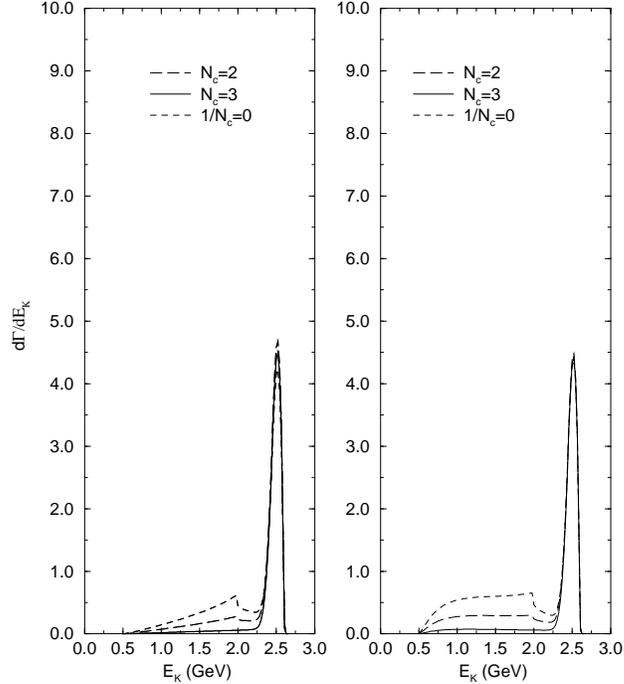}}
\caption{Predicted rate for $B^-\rightarrow K^- X$ as a function of the
kaon energy. Three curves in 
each figure are shown for $N_c=2, 3, \infty$ and provide
an estimate of the theoretical uncertainty from the
factorization hypothesis.
In (a) and (b) different sets of form factors are then compared
 in order to determine
the sensitivity of the predicted rate
to the choice of a form factor model. 
 The vertical scale in the plots  is multiplied by
$(G_F/\protect\sqrt{2})^2 \times 10^{-6} $.}
\label{bkx_rate}
\end{figure}

Using the form of $M_1$ and $M_2$ given above one can write the partial
rate asymmetries as
\ber
{a}^{ij} &= &
\frac{2Im({V_u}^* V_c)Im({\Delta_{ut}^{i}}{\Delta_{ct}^j}^{*})}
    { \sum_{ij} \left( |V_u|^2|{\Delta_{ut}^{i}}{\Delta_{ut}^j}^{*}|  +
     |V_c|^2 |{\Delta_{ct}^{i}}{\Delta_{ct}^j}^{*}|  + 
2Re({V_u}^* V_c)Re({\Delta_{ut}^{i}}{\Delta_{ct}^j}^{*}) \right) }\
\eer
where $i, j =1, 2$. The net asymmetry, $a$, is given by the sum 
\ber
a = a^{1 1} + a^{2 2} + a^{1 2}+ a^{21}
\eer
From the values of the Wilson coefficients $c_1$ and $c_2$ it can be shown
that the contribution to the asymmetry due to the interference of the tree
and penguin amplitudes is suppressed in ${a}^{11}$. This coupled
with the fact that the gluon momentum $q^2$ is varying for $M_1$ while
it is more or less fixed for $M_2$ can lead to a larger value for
${a}^{2 2}$ compared to ${a}^{1 1}$ and ${a}^{1 2}$. 
It should be pointed out that the
interference term between $M_1$ and $M_2$ can be important when
calculating the partial rate asymmetries.

\section{Results and Discussion of Theoretical Uncertainties}
\begin{figure}[htb]
\centerline{\epsfysize 3.4 truein \epsfbox{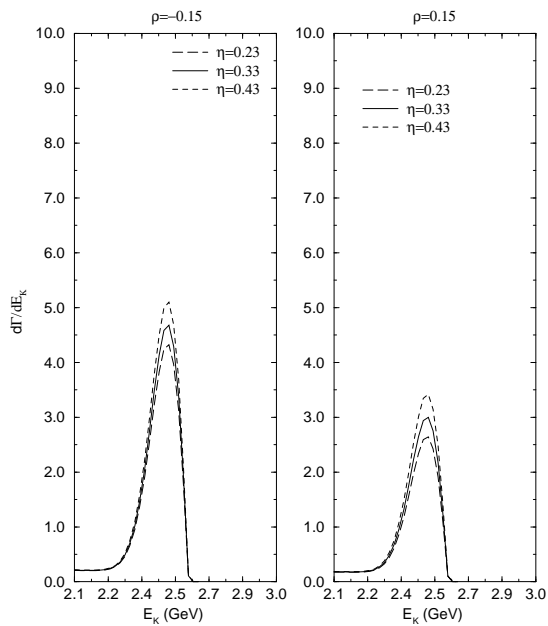}}
\caption{Sensitivity of
the predicted rate for $B^-\to K^- X$ to $\rho, \eta$.
The three curves 
indicate the sensitivity of the rate to the value of Wolfenstein
parameter $ ~\eta$ for fixed values of $\rho$}
\label{bkx_rhoeta}
\end{figure}

\begin{figure}[htb]
\centerline{\epsfysize 3.4 truein \epsfbox{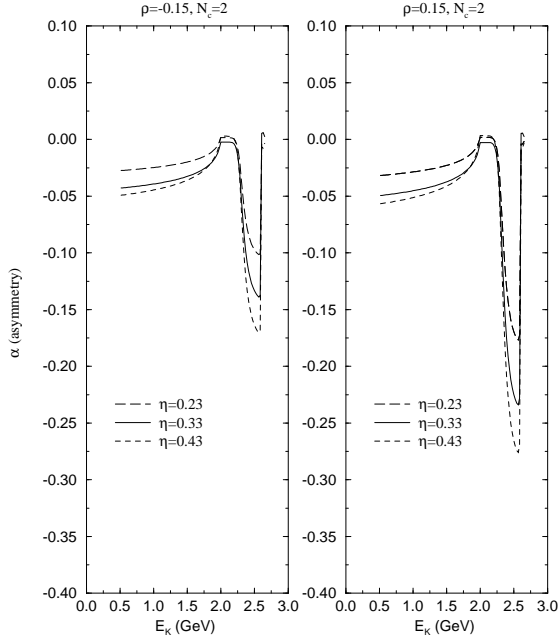}}
\caption{Sensitivity of the
 asymmetry for $B^-\to K^- X$ to the Wolfenstein parameter
$\eta$ for two fixed values of $\rho$. The three curves
indicate the sensitivity of the asymmetry as a function of kaon energy.}
\label{bkxasym_rhoeta}
\end{figure}

\begin{figure}[htb]
\centerline{\epsfysize 2.8 truein \epsfbox{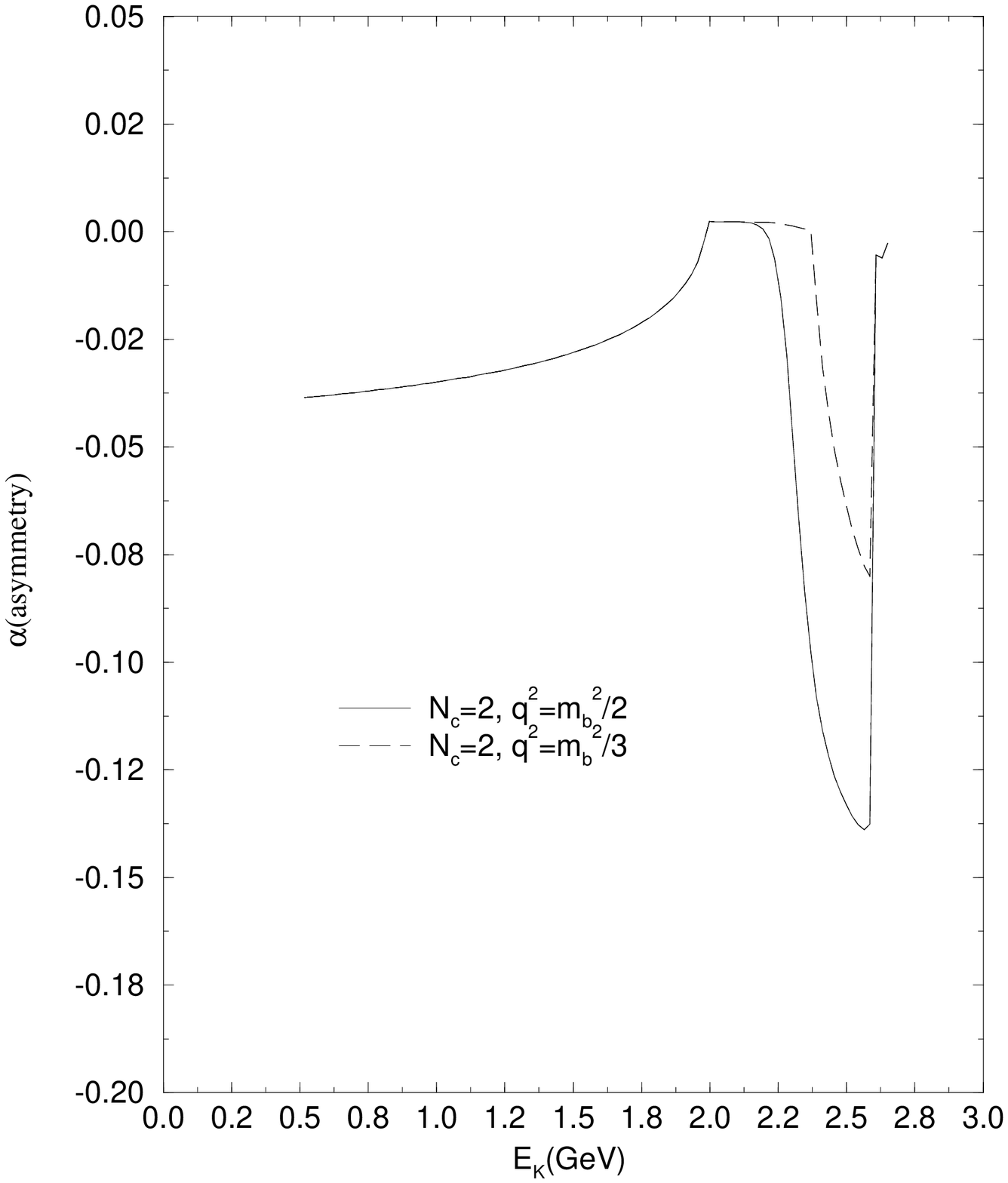} ~~~~
\epsfysize 2.8 truein \epsfbox{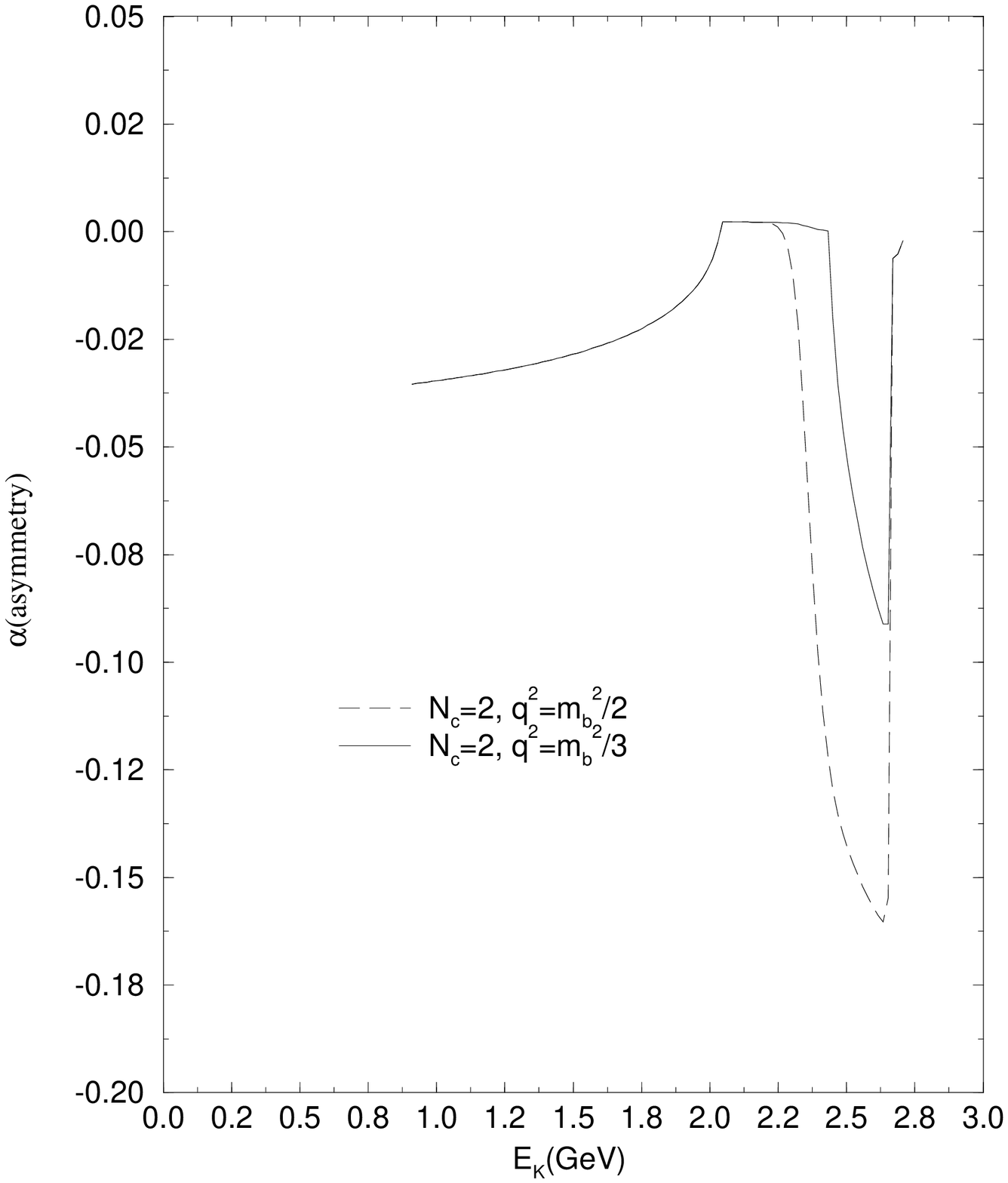}}
\caption{Predicted asymmetries for 
$B^-\to K^- X$ and $B^-\to K^{*-} X$ as a function of the
kaon energy. The value of $N_c=2$ is fixed. 
The two sets of curves
indicate the sensitivity of the asymmetry to the values of $q^2$ for
the gluon in the internal two body diagram.}
\label{bkx_qsquared}
\end{figure}

\begin{figure}[htb]
\centerline{\epsfysize 2.8 truein \epsfbox{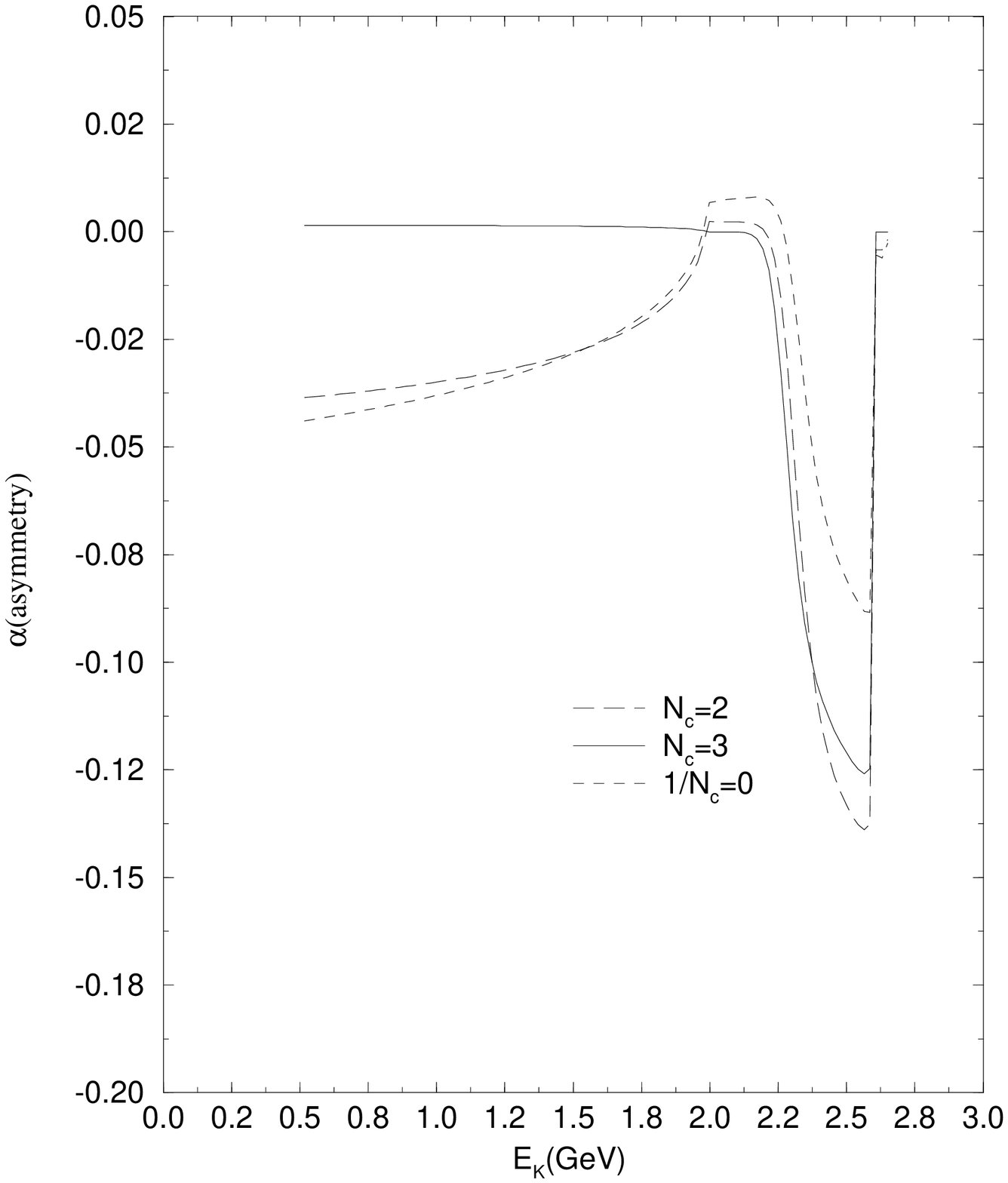} ~~~~
\epsfysize 2.8 truein \epsfbox{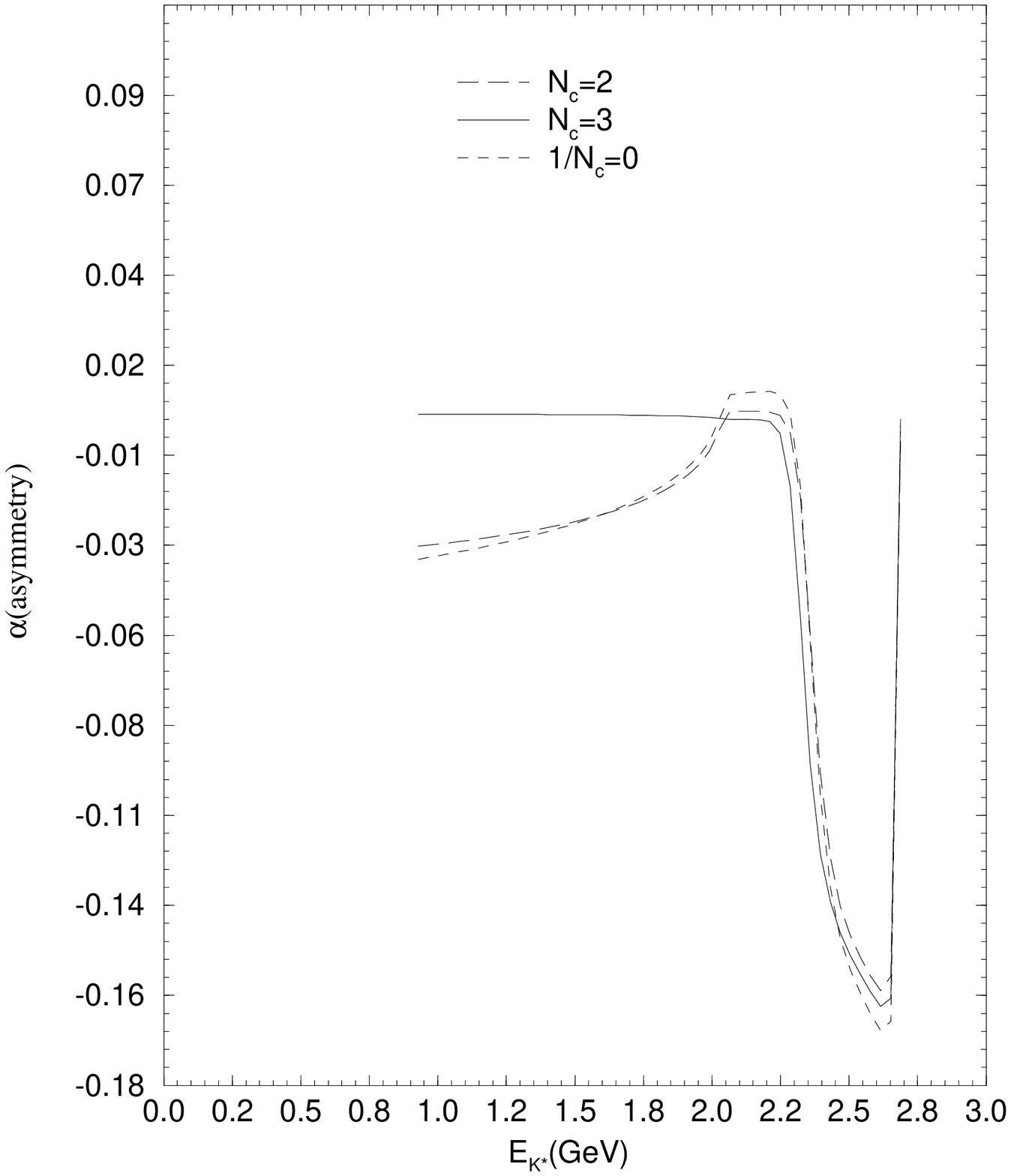}}
\caption{Predicted asymmetries for 
$B^-\to K^- X$ and $B^-\to K^{*-} X$ as a function of the
kaon energy. The three sets of curves
indicate the sensitivity of the asymmetry to the value of $N_c$.
The values $N_c=2, 3, \infty$ are considered.}
\label{bkxasym_Nc}
\end{figure}

\begin{figure}[htb]
\centerline{\epsfysize 3.4 truein \epsfbox{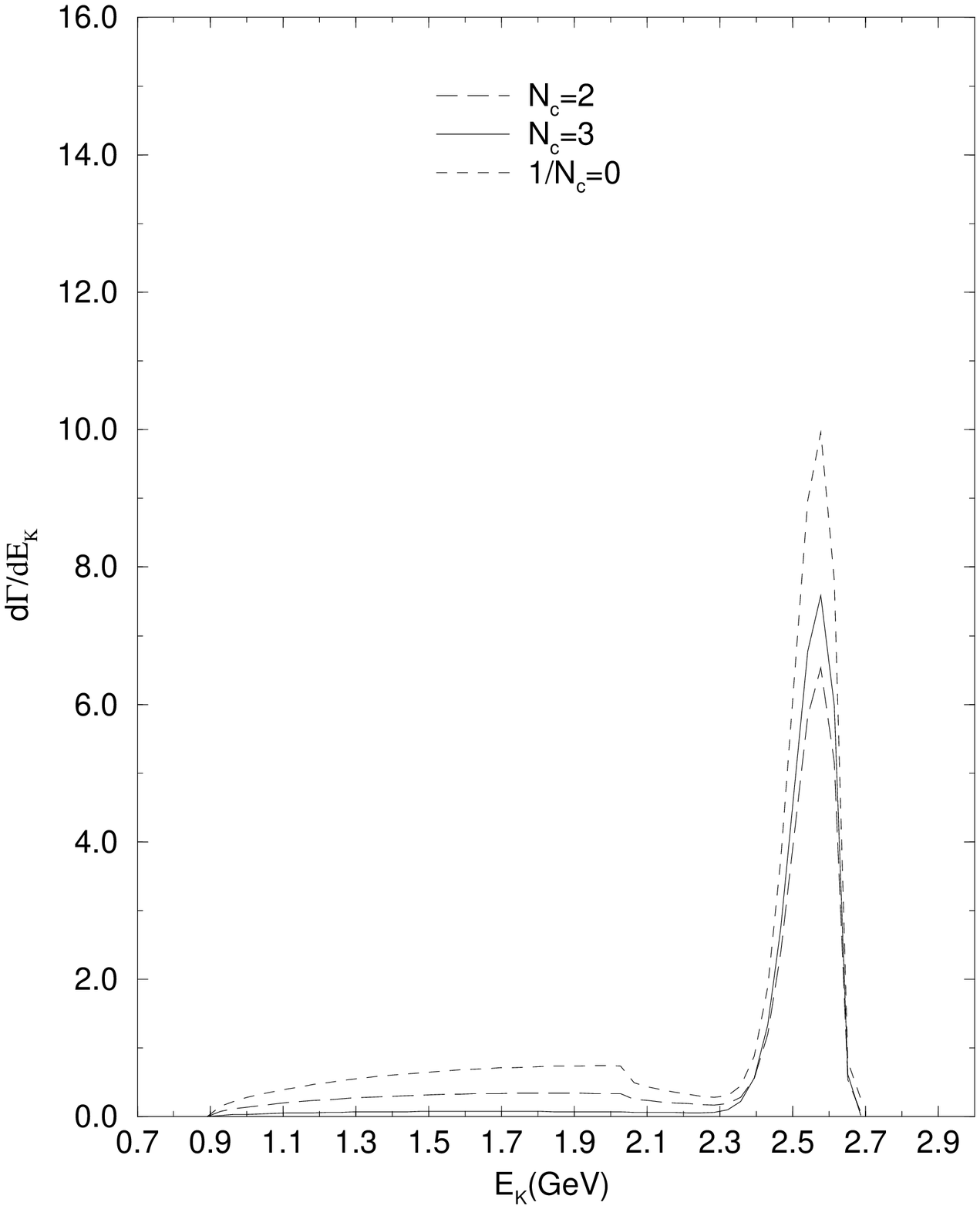}}
\caption{Predicted rate for $B^-\to K^{*-} X$ as a function of the
kaon energy. 
The three curves
indicate the sensitivity of the rate to the value of $N_c$.
The values $N_c=2, 3, \infty$ are considered.}
\label{bkstxrate_Nc}
\end{figure}

\begin{figure}[htb]
\centerline{\epsfysize 3.4 truein \epsfbox{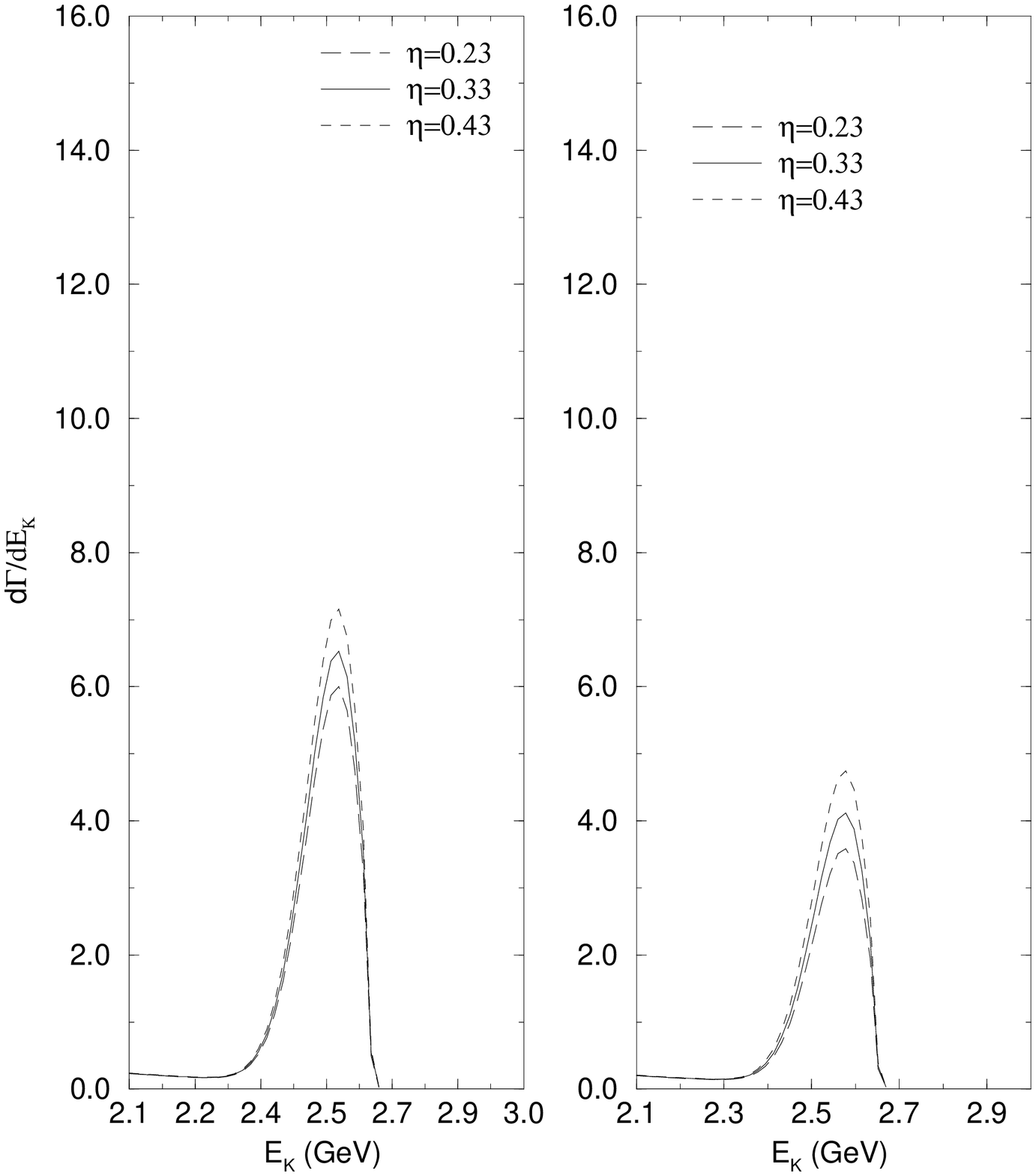}}
\caption{Predicted rate for $B^-\to K^{*-} X$ as a function of the
kaon energy. 
The three curves
indicate the sensitivity of the rate to the values of the Wolfenstein
parameters $\rho, ~\eta$. Only the signal region is shown in the figures.}
\label{bkstxrate_rhoeta}
\end{figure}

\begin{figure}[htb]
\centerline{\epsfysize 3.4 truein \epsfbox{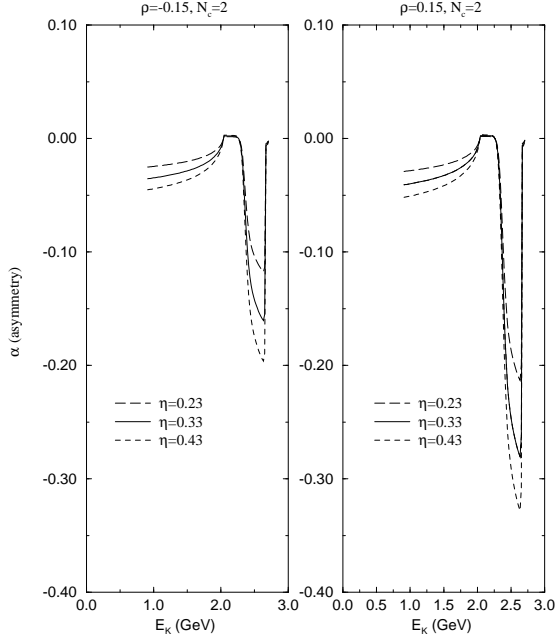}}
\caption{Predicted asymmetry for $B^-\to K^{*-} X$ as a function of the
kaon energy. 
The three curves
indicate the sensitivity of the asymmetry to the values of the Wolfenstein
parameters $\rho, ~\eta$.}
\label{bkstxasym_rhoeta}
\end{figure}

\begin{figure}[htb]
\centerline{\epsfysize 2.8 truein \epsfbox{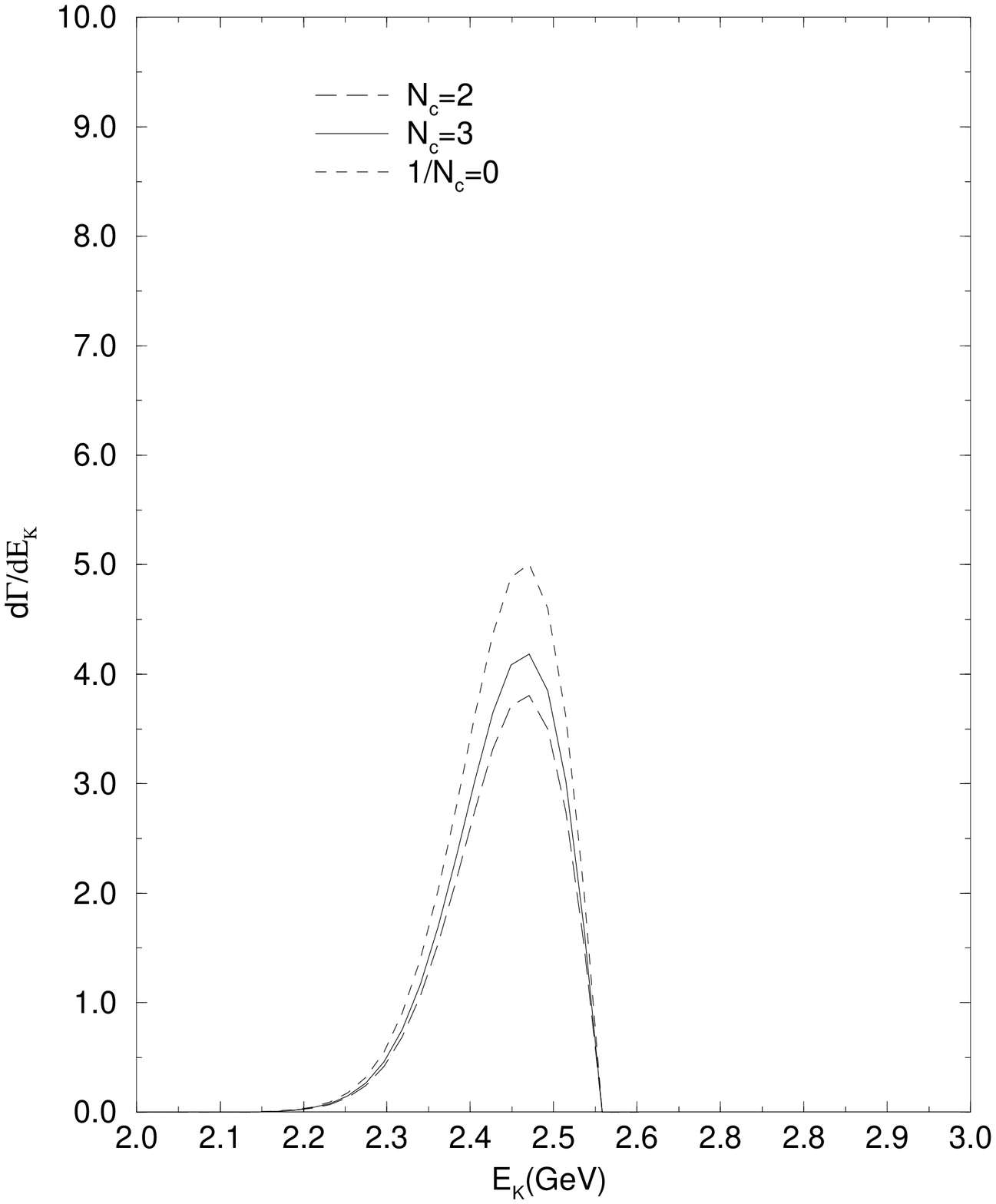} ~~~~
\epsfysize 2.8 truein \epsfbox{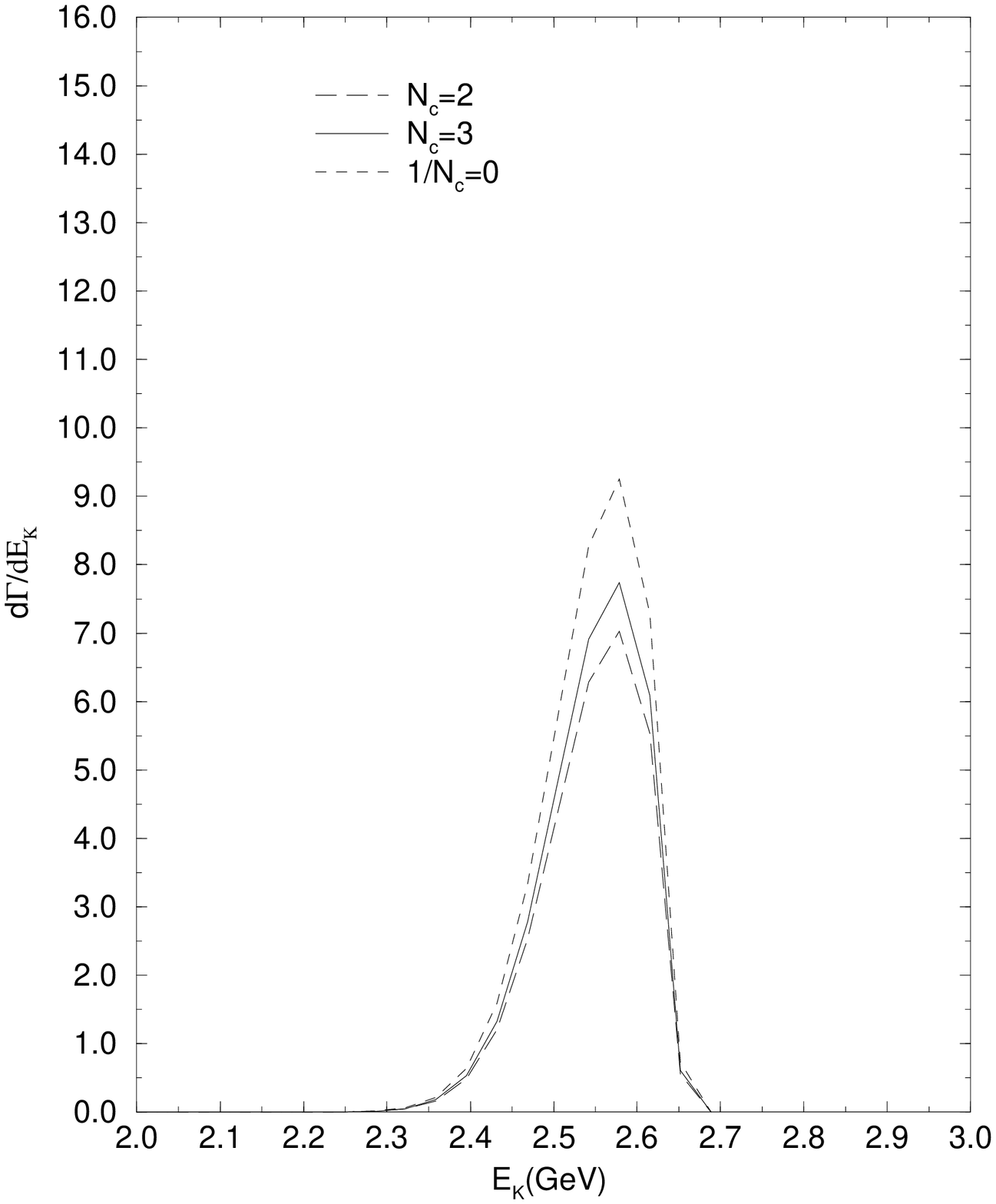}}
\caption{Predicted rate for 
$\bar{B}^0\to K^{-} X$  and $\bar{B}^0\to K^{*-} X$ 
as a function of the kaon energy. 
The three sets of curves
indicate the sensitivity of the rate to the value of $N_c$.
The values $N_c=2, 3, \infty$ are considered.}
\label{bkxrate_b0_Nc}
\end{figure}

\begin{figure}[htb]
\centerline{\epsfysize 3.4 truein \epsfbox{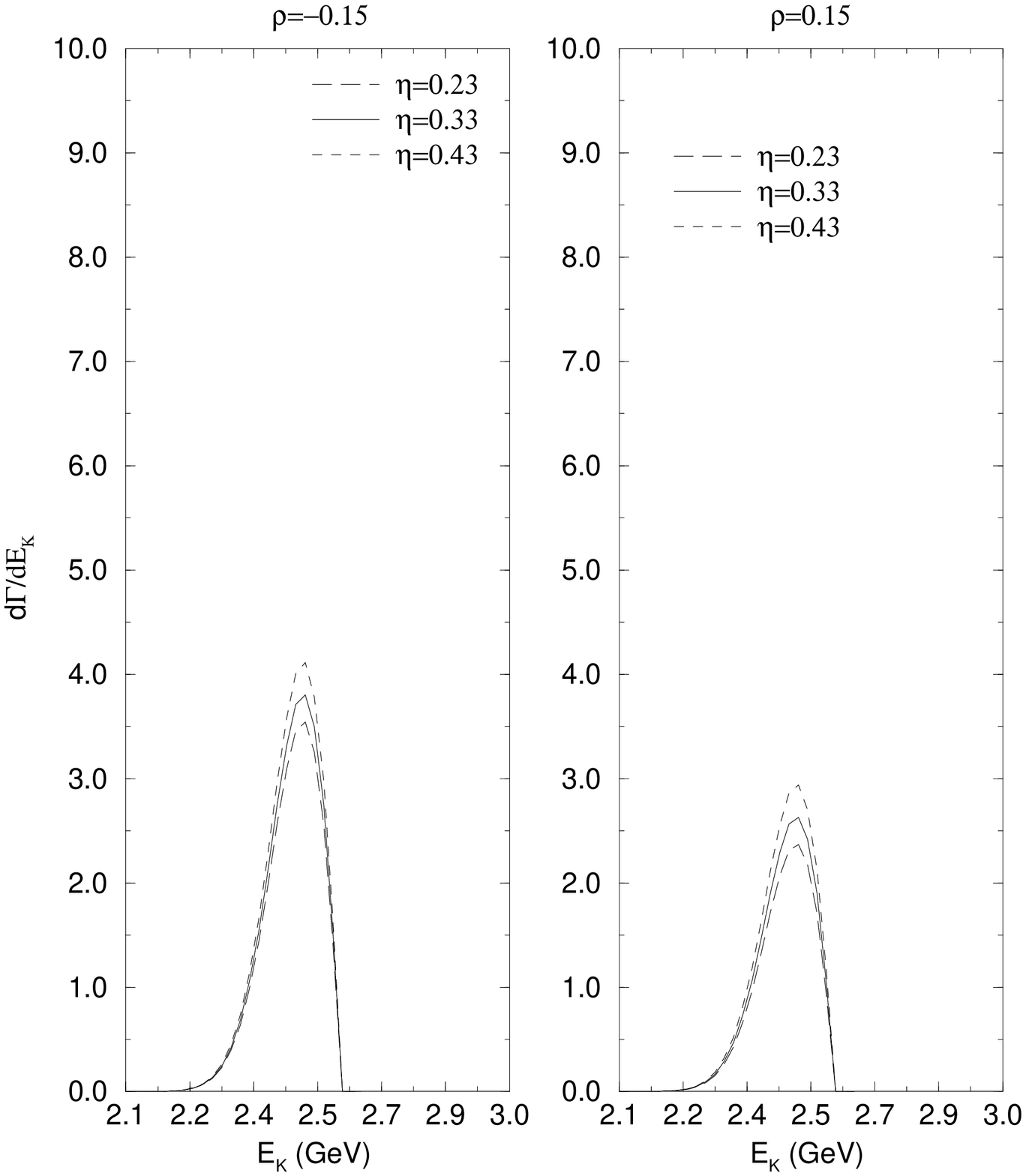}}
\caption{Predicted rate for $\bar{B}^0\to K^{-} X$ as a function of the
kaon energy. 
The three curves
indicate the sensitivity of the rate to the values of the Wolfenstein
parameters $\rho, ~\eta$.}
\label{bkxrate_b0_rhoeta}
\end{figure}

\begin{figure}[htb]
\centerline{\epsfysize 3.4 truein \epsfbox{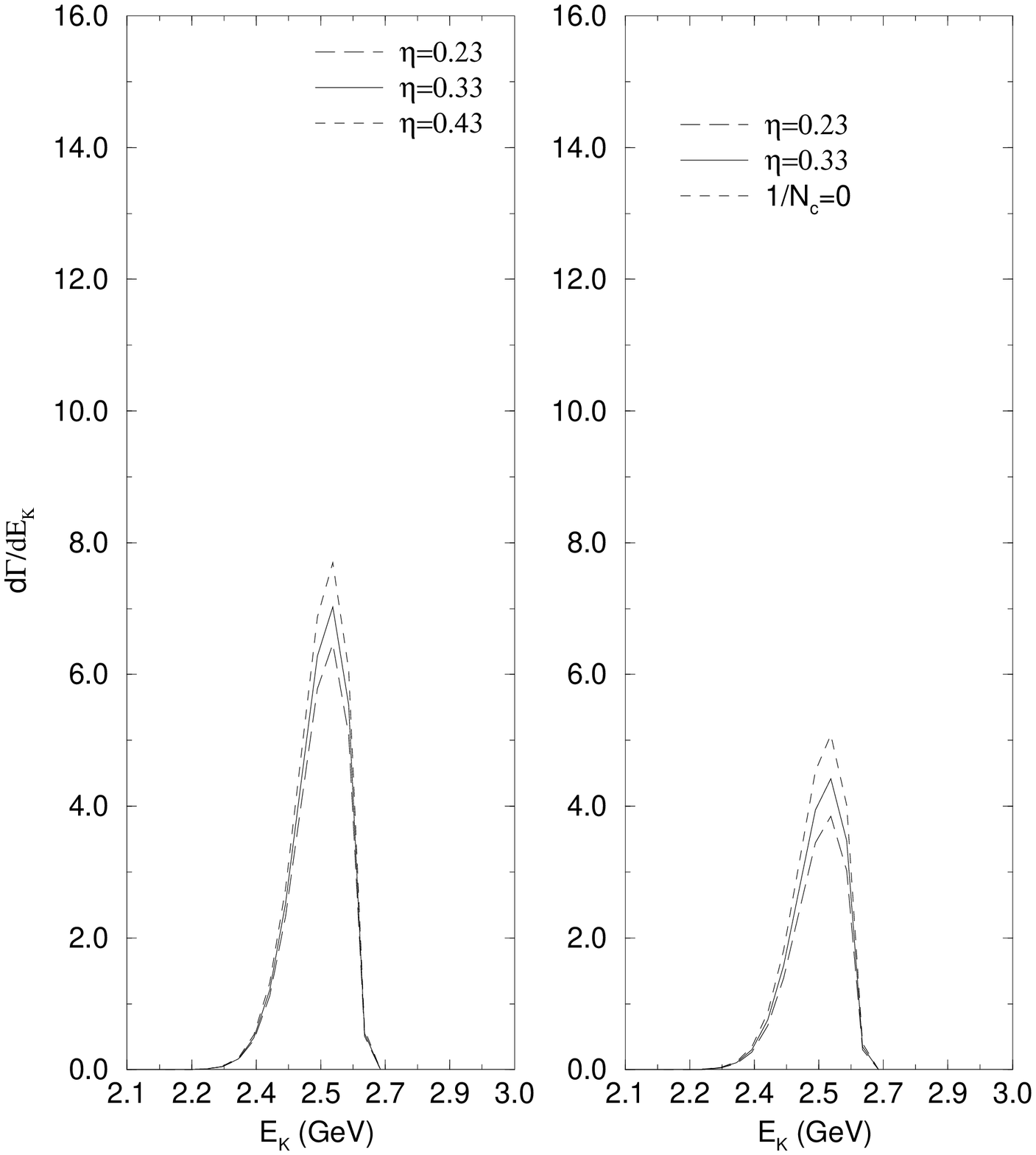}}
\caption{Predicted rate for $\bar{B}^0\to K^{*-} X$ as a function of the
kaon energy. 
The three curves
indicate the sensitivity of the rate to the values of the Wolfenstein
parameters $\rho, ~\eta$.}
\label{bkstxrate_b0_rhoeta}
\end{figure}

\begin{figure}[htb]
\centerline{\epsfxsize 2.3 truein \epsfbox{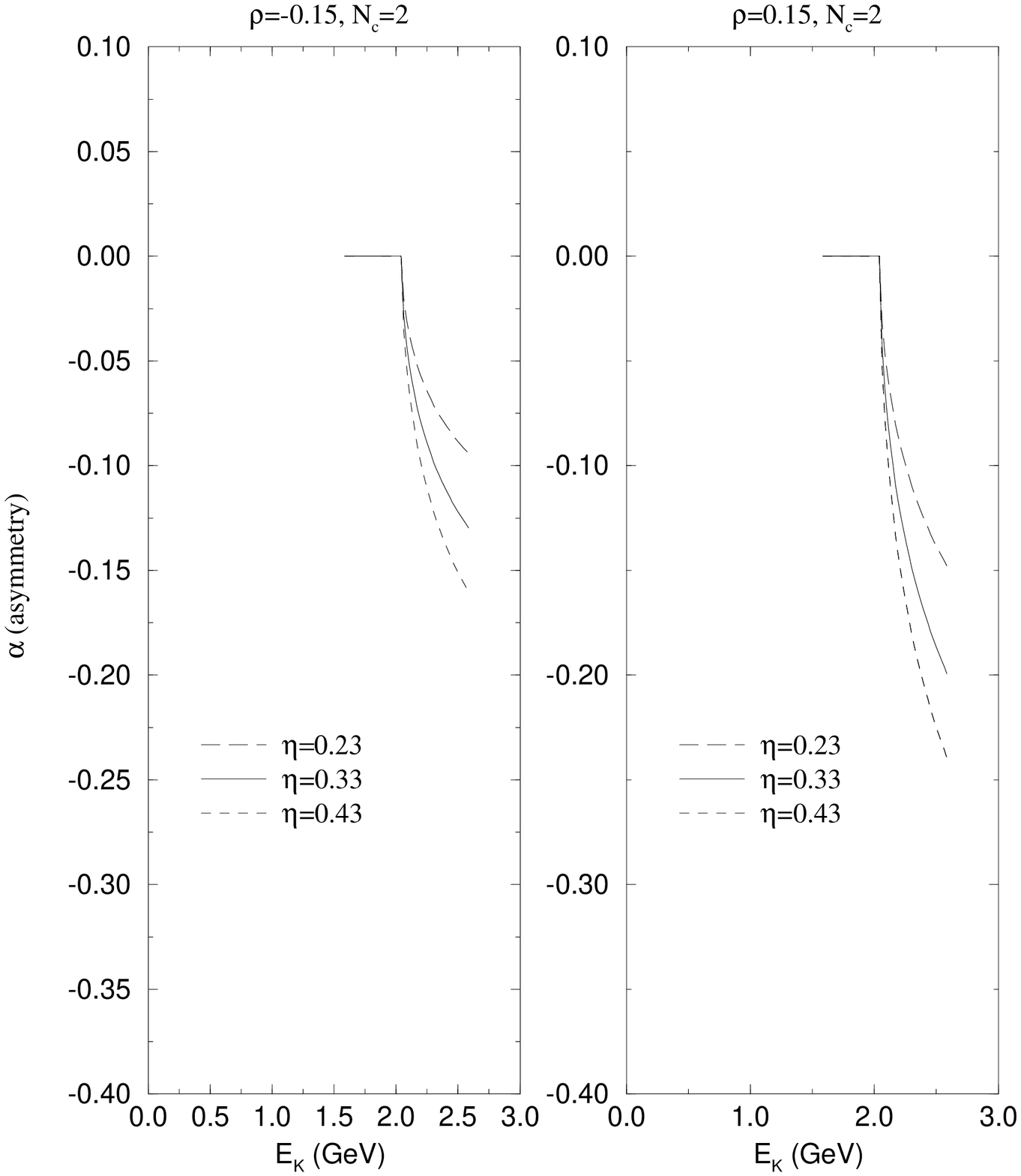} ~~~~~~~~~~~~~~
\epsfxsize 2.3 truein \epsfbox{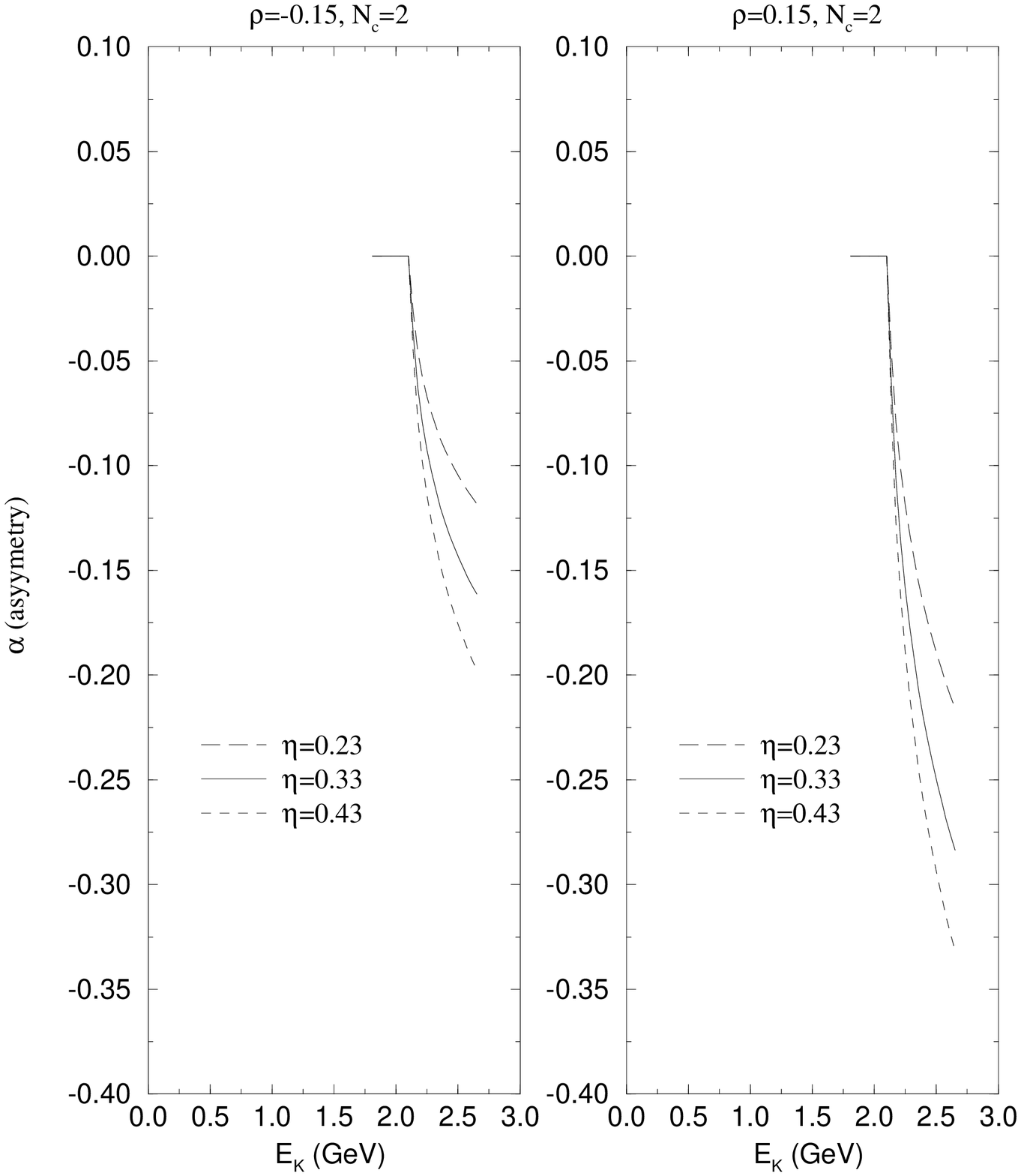}}
\caption{Predicted asymmetries for $\bar{B}^0\to K^{-} X$ 
and $\bar{B}^0\to K^{*-} X$ as a function of the
kaon energy. The three curves in each figure
indicate the sensitivity of the rate to the values of the Wolfenstein
parameters $\rho, ~\eta$.}
\label{bkxasym_b0_rhoeta}
\end{figure}

\begin{figure}[htb]
\centerline{\epsfysize 2.8 truein \epsfbox{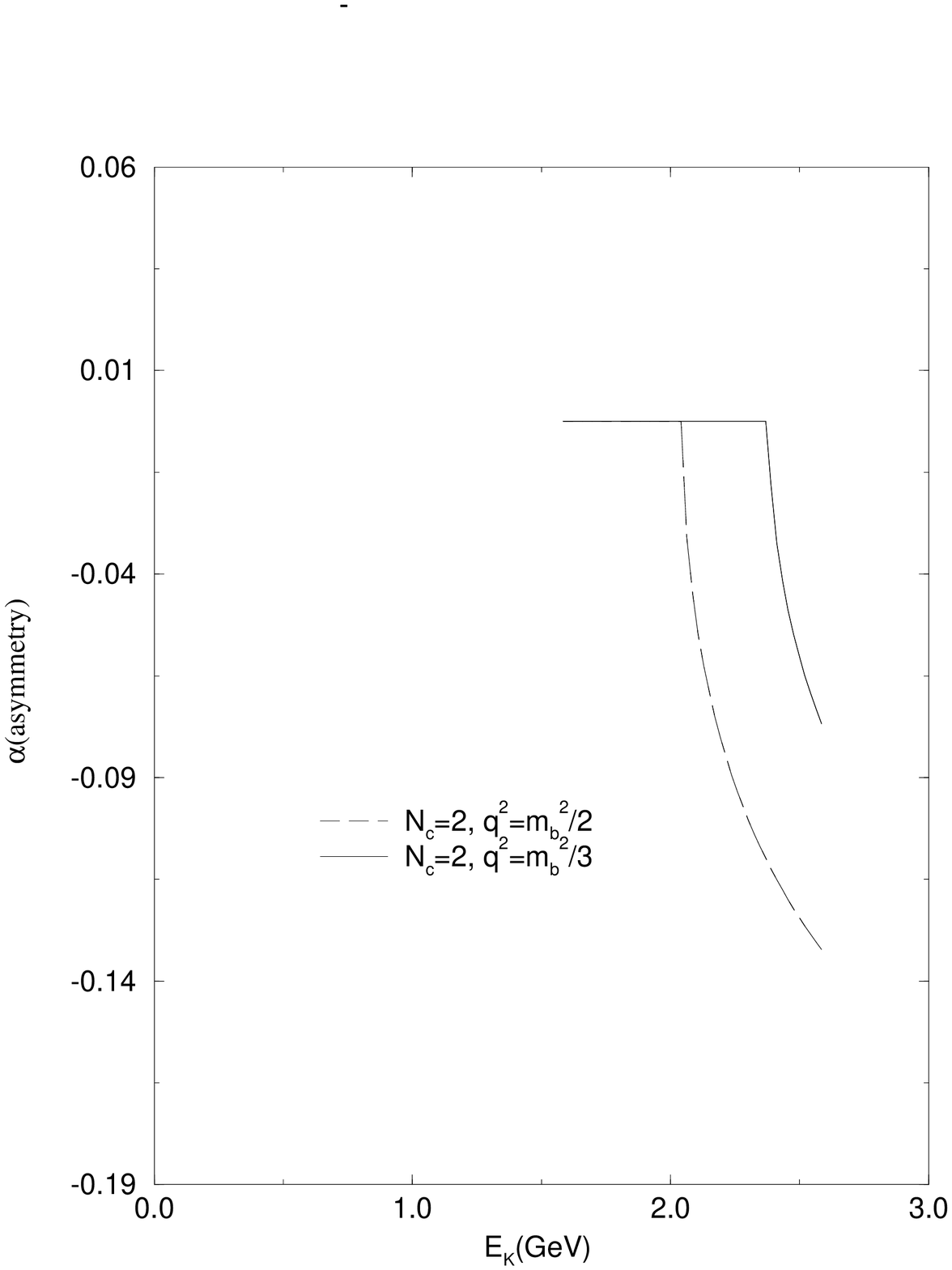} ~~~~
\epsfysize 2.8 truein \epsfbox{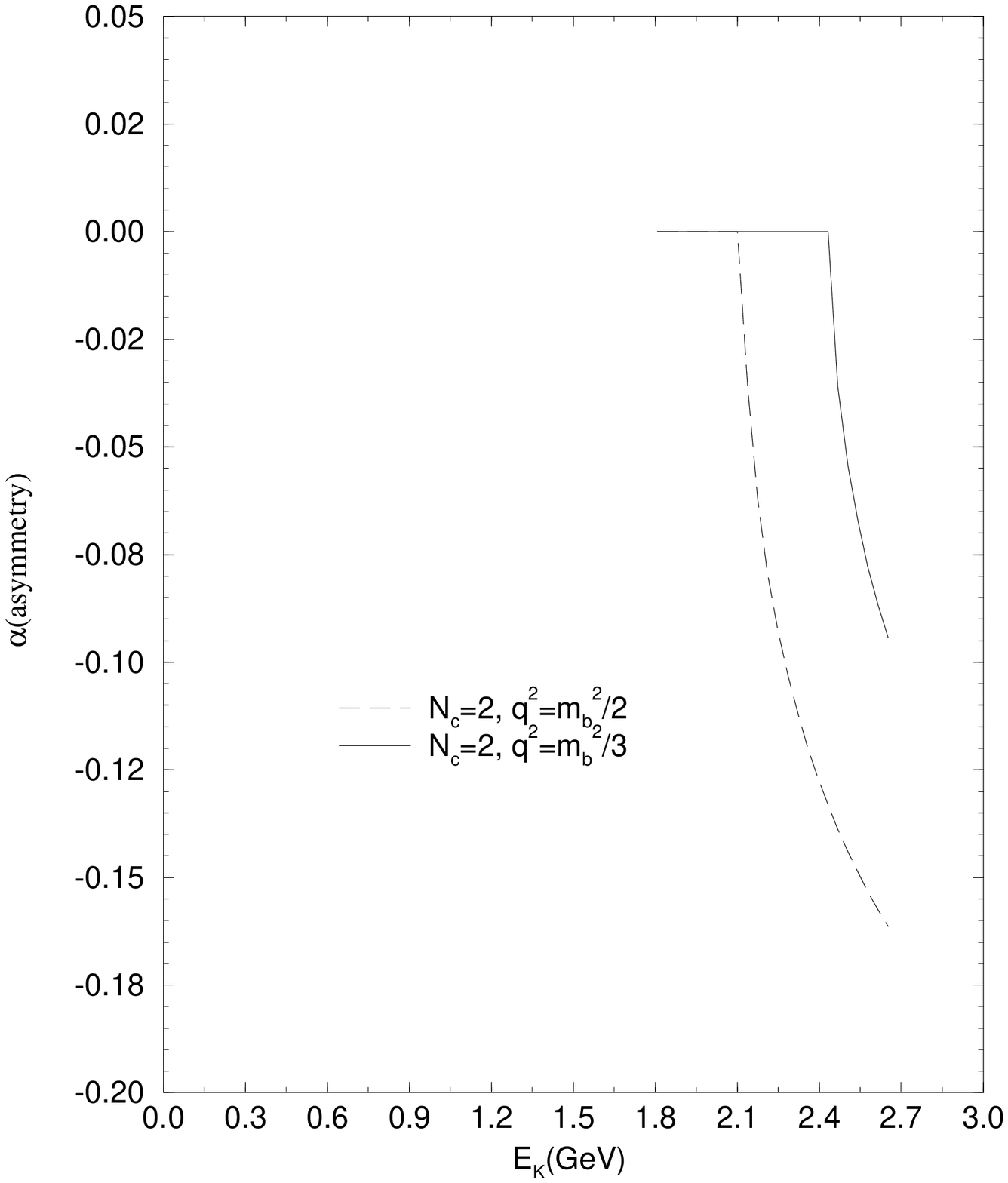}}
\caption{Predicted asymmetries for 
$\bar{B}^0\to K^{-} X$ and $\bar{B}^0\to K^{*-} X$ as a function of the
kaon energy. The two sets of curves
indicate the sensitivity of the asymmetry to the values of $q^2$ for
the gluon in the internal two body diagram.}
\label{bkxasym_b0_qsquared}
\end{figure}

 In this section we discuss the results of our calculations which are
shown graphically in Figs. \ref{bkx_rate} -~\ref{bkxasym_b0_qsquared}. 
We find that there can be significant asymmetries in $B \to K (K^*) X$
decays especially in the region  $E_K> 2 $ GeV which is also the region
where an experimental  signal for such decays can be isolated.
The branching ratios are of order $O(10^{-4})$ 
which are within reach for future $B$ factories.

The contribution of the amplitude with the top quark in the loop
accounts for 60-75\% of the inclusive branching fraction. However,
since the top quark amplitude is large and has no absorptive part
in contrast to the c quark amplitude, 
the top quark contribution reduces the net CP asymmetry from 30-50\% to
about 10\%. This calculation includes the contribution from
electroweak penguins. We find that the electroweak penguin 
contributions increase the decay rates by 10-20\% but reduce
the overall asymmetry by 20-30\%.

Let us now identify the main sources of uncertainties in our
calculation. These are the use of the factorization approximation,
the choice of a form factor model, the
choice of $q^2$ for the gluon momentum in the G function in Eq. (4)
 for the ``internal'' two body
diagrams, and the choice of a model for the $B$-meson. We now discuss
the sensitivity of the results for decay rates and asymmetries
 to these theoretical uncertainties.

We have used the factorization approximation.
The factorization approximation is expected to be valid in our
calculations as we have argued at the beginning of Section 2.
To take into 
account corrections to this approximation we allowed the number
of colors to be a free parameter. In our calculation we consider 
the cases $N_c = 2, ~3, ~\infty$ although the analysis of exclusive
two body $B$ decays suggests that $N_c \sim 2$.
In Fig.~\ref{bkx_rate}
we show the decay distribution for $B^-\rightarrow K^- X$
where $X$ does not contain any strange particles.
In the region of interest to experiment (i.e. $E_K> 2.0$ GeV) the decay
distribution has only a modest dependence on $N_c$.

The second source of uncertainty is the choice of form
factors used to describe the $B\rightarrow K(K^*)$ transitions. As mentioned
earlier we use the form factors given in Ref.~\cite {Narduli}. In
Ref~\cite{Narduli}
a monopole $k^2$ dependence for the form factors is chosen and the form
factor at $k^2 =0$ for the D decays is fixed from semileptonic $D\to K$ decays.
The form factors are then scaled  to the case of $B \rightarrow K$ 
decay using heavy quark effective theory
(HQET). The primary effect of choosing a different set of form
factors in our calculations
 is to change the decay distribution of the ``three body
distribution'' which is governed by the matrix element $M_1$.
Figures. 3(a), 3(b) show
the effect of choosing two sets of form factors 
from Ref~\cite{Narduli} and Ref~\cite{BSW1}
with a different $k^2$ behavior of the form factors. For the
energy region $E_K> 2$ GeV the form factor effects are
negligible for the decay distributions as well as the asymmetries.
For the remainder of the discussion and 
in the rest of the figures we will only show the results with form factors 
from Deandrea et al.~\cite{Narduli}.

 The asymmetries are sensitive to the values of 
the Wolfenstein parameters $\rho$ and $\eta $. The 
existing constraints on the  values of $\rho$ and
$\eta$ come from measurements of $|V_{ub}|/|V_{cb}|$, $\epsilon_K$ in the
K system and $\Delta M_{B_d}$. (See  Ref.~\cite{Ali1} for a recent review).
In our calculation we will use $f_B =170$ MeV  and choose
$(\rho=-0.15,\eta=0.33)$. 
To determine the dependence of our results on
$\eta$ we will also consider three sets of representative values,
 $(\rho=-0.15,\eta=0.23)$, 
$(\rho=-0.15,\eta=0.33)$ and $(\rho=-0.15,\eta=0.43)$. 
We will also
consider the set of $\eta$ values with $\rho=0.15$. The dependence
of the rates on $\rho, \eta$ is shown in
Fig.~\ref{bkx_rhoeta}, while the sensitivity of 
the asymmetry to these parameters is shown 
in Fig.~\ref{bkxasym_rhoeta}. For fixed $\rho$, the asymmetry 
increases monotonically as $\eta$ increases. The results suggest
that measurement of asymmetries in inclusive decays will give useful
information on $\eta$ once the size of the theoretical
uncertainties is reduced.

There are several other sources of uncertainty. These are: (1) the
choice of $q^2$ for the gluon momentum in the G function in Eq. (4)
 for the ``internal'' two body
diagrams and (2) the choice of a model for the 
wavefunction of the $B$-meson. The $q^2$
variation causes a small change in the decay distribution but a fairly
significant change in the asymmetries. This uncertainty is illustrated
by comparison of the two curves in Figs.~\ref{bkx_qsquared}, 
\ref{bkxasym_b0_qsquared}. 

The choice of
the value of the charm quark mass $m_c$, 
is also a source of uncertainty for the asymmetry. We have taken
$m_c=1.35$ GeV for this calculation. Increasing the charm quark
mass to $m_c=1.6$ GeV does not significantly modify the 
decay rates but reduces the asymmetry by about 30\%. 
Since the Wilson
coefficients are calculated to next to leading order, they have little
sensitivity to the renormalization scale. However, 
the G function
which enters into the calculation has a stronger dependence on
renormalization scale. Varying $\mu^2$ from $m_b^2/2$ to $2 m_b^2$ changes 
both the asymmetry and the decay rates by $\pm 10\%$. A lower
renormalization scale corresponds to a larger asymmetry.  
The model of  the $B$ meson  wavefunction, used to take into account
the Fermi motion of the quarks, is yet another source of uncertainty. 
We have used the model of Ali {\it et al.} which has been previously
used for the $B\rightarrow X_s \gamma$ case\cite{ABS}. 
Different models give somewhat different results for the decay distributions
while the asymmetries are insensitive 
to the choice of model for the $B$ meson wavefunction.

In Fig.~\ref{bkstxrate_Nc}
we show the decay distribution for $K^*$ in the final
state while in Figs. \ref{bkxasym_b0_rhoeta},
~\ref{bkstxrate_b0_rhoeta} we show the variation of the decay
distribution and asymmetry with different sets of $\rho,\eta$ for $N_c=2$.
In Fig. 11
we show the asymmetry for several values of $N_c$ using
the form factors from Ref\cite{Narduli} and $q^2=m_b^2/2.0$. 
A similar variation of the asymmetry with $q^2$ as
calculated for the $K$ in the final state is also observed in this case
as shown in Fig. 10.
 
Turning to $B^0$ decays, only the $M_2$ part
of the matrix element contributes. In Fig.~\ref{bkxrate_b0_Nc}
we show the decay distribution for various  $N_c$ values
using the form factors from Ref\cite{Narduli} and
$q^2=m_b^2/2.0$. 
In Fig.~\ref{bkxrate_b0_rhoeta}
we show the decay distributions of $\bar{B}^0\to K^-$
for representative values of $(\rho,\eta)$. In Figs.~\ref{bkxasym_b0_rhoeta},
\ref{bkxasym_b0_qsquared}
we show the asymmetries as we vary $(\rho,\eta)$ and $q^2$. The
variation of the asymmetries with $N_c$ is negligible in this case.

In Table.~\ref{Tb_integrated}
we give the branching fractions and the integrated
asymmetries for the inclusive decays for different N, $q^2=m_b^2/2$,
$f_B=170$ MeV, $\rho=-0.15, \eta=0.33 $ and 
the form factors from Ref.~\cite{Narduli}. For the charged $B$ decays we
also show the decay rates and asymmetries for 
$E_K>2 $ ($2.1$) GeV as that is the region of the signal.
%

%
\begin{table}
\caption{Integrated decay rates and asymmetries for $B\to K^{(*)} X$ Decay}
\begin{center}
\begin{tabular}{ccc}
 Process &  Branching ratio ($1.65\times 10^{-4}$) &  
Integrated asymmetry  \\ \hline
    &  & \\
$B^-\rightarrow K^- X$ & $1.02,0.79,1.20$ &$-0.10,-0.11,-0.050$\\

$B^-\rightarrow K^- X (E_K>2.1 \rm{GeV})$ & $0.81,0.74,0.77$
&$-0.12,-0.12,-0.07$\\

${\overline B^0}\rightarrow K^- X$ 
$(E_K>2.1 \rm{GeV})$ & $0.6,0.7,0.8$ & $-0.12,-0.12,-0.13$\\
%
     &  & \\
$B^-\rightarrow K^{*-} X$ & $1.37,1.24,2.30$ & $-0.11,-0.14,-0.11$\\

$B^-\rightarrow K^{*-} X (E_{K^*} >2.1 \rm{GeV})$
 & $1.05,1.16,1.67$ &$-0.14,-0.15,-0.14$\\

${\overline B^0}\rightarrow K^{*-}X$ $(E_{K^*} >2.1 \rm{GeV})$ 
& $1.05,1.16,1.39$ & $-0.15,-0.15,-0.16$\\
\end{tabular}
\end{center}
\label{Tb_integrated}
\end{table}

The above figures
show that there can be significant asymmetries in $B \to K^{(*)} X$
decays, especially in the region  $E_K> 2 $ GeV which is the region
of experimental sensitivity 
 for such decays. As already mentioned, our calculation is
not free of theoretical uncertainties. Two strong assumptions used in
our calculation are the use of quark level strong phases for the FSI
phases at the hadronic level and the choice of the
 value of the gluon momentum $q^2$ in the two body decays. Other
uncertainties from the use of different heavy to light
form factors, the use of factorization, the model of the
$B$ meson wavefunction, the value of the charm quark
mass and the choice of the renormalization scale $\mu$ 
have smaller effects on the asymmetries.

The use of quark level strong phases at the hadronic level
neglects the possibility of soft FSI. This neglect may be a better
approximation in our inclusive case as opposed to exclusive
modes.

\section{Comparison with Previous Calculations}

Inclusive direct CP violating asymmetries
were calculated by Gerard and Hou\cite{HouW}. They obtained asymmetries
much smaller than the results given here. The differences between
these calculations and our results can be understood as follows.
In the Gerard-Hou calculation only
the three body quark level process corresponding to $b\to s q \bar{q}$ 
(i.e. $M_1$ in equation (1)) was considered and the limit $N_c\to
\infty$ was taken. With these conditions we obtain an asymmetry of
$-1.7\%$ which then agrees with their result. 
We obtain large asymmetries only in the kinematic
regime dominated by the two-body process $b\to K ~u$. These asymmetries
agree qualitatively with the asymmetry for exclusive modes such as
$B^-\to K^-\pi^0$ found in several recent calculations 
in both sign and magnitude.
In the kinematic region of $B^-\to K^-\pi^0$
we find an asymmetry of $-(8-14)\%$ to be compared with 
and $-(2-8)\%$ found by Kramer, Palmer, and Simma\cite{Kramer} 
and $-(3-9)\%$ found by
Kamal and Luo\cite{Kamal} for the exclusive mode $B^-\to K^-\pi^0$.

We have also verified that the asymmetry in the process
$b\to s c \bar{c}$ has the opposite sign to the asymmetry in 
the sum of $b\to s u \bar{u}$, $b\to s d\bar{d}$, $b\to s  s \bar{s}$
(as does the asymmetry in $B\to D\bar{D}$) as required by CPT
to ensure the cancellation between all modes.

\section{Conclusion}

We find significant direct CP violation in the inclusive decay
$B\to K^{-} X$ and $B\to K^{*-} X$ for $2.7>E_{K^{(*)}}>2.0$ GeV.
The branching fractions are in the $10^{-4}$ range and the CP
asymmetries may be sizeable. These asymmetries should be observable
at future $B$ factories and could be used to rule out the superweak
class of models.

\section{Acknowledgements}

We thank K. Berkelman, W-S. Hou, A. Kagan, L. Wolfenstein, and
H. Yamamoto for useful discussions and comments. This work was
supported by the Australian Research Council and by the
United States Department of Energy under contracts
DE-FG 02-92ER10730 and DE-FG 03-94ER40833. X.-G. He thanks
the University of Iowa and the University of Oregon for hospitality
during the preparation of this work. 

\clearpage
\appendix
\section{}

For $B^- \rightarrow K^-X$ decay
\ber
	{A^1}_{\mu\nu} H^{\mu\nu} & =& 
|g_+|^2 D(p_B, p_B) + |g_-|^2 D(p_K, p_K)
				+ 2 Re (g^{+} g_-^*) D(p_B, p_K)\\
{A^2}_{\mu\nu} H^{\mu\nu} & =& 0\
\eer
where
\ber
	g_+ & = & f_+ + f_- \nonumber\\
	g_- & = & f_+ - f_-
\eer
and we have defined
\be
	D(A,B) = A \cdot p_1 \, B \cdot p_2 + B \cdot p_1 \, A \cdot p_2 
		- p_1 \cdot p_2 \, A \cdot B
\ee
\be
	g_{\mu\nu} H^{\mu\nu} = |g_+|^2 m_B^2 + |g_-|^2 m_K^2
				+ 2 Re (g_ + g_-^*) \, p_B \cdot p_K
\ee

For $B^- \rightarrow K^{*-} X$ decays (with $ p=p_B+p_{K^*}, q=p_B-p_{K^*}$),
\be
	H^{\mu\nu} {A^1}_{\mu\nu} = \sum T_i
\ee
\ber
	T_1 & = &  b_1^2 \{ -2 p_1 \cdot p_2 (m_B^2 m_{K^*}^2 
		- (p_B \cdot p_{K^*})^2 ) \nonumber \\
	    & &	+ 2 p_B \cdot p_{K^*} D(p_B, p_{K^*})
		- {m_{K^*}}^2 D(p_B, p_B) 
		- m_B^2 D(p_{K^*}, p_{K^*}) \} \nonumber\\
	T_2 & = & b_2^2 \left[2 p_1 \cdot p_2 
		+ \frac{1}{m_{K^*}^2} D(p_{K^*}, p_{K^*}) \right]\nonumber \\
	T_3 & = & b_3^2 F D(p, q) \nonumber\\
	T_4 & = & b_4^2 F D(p, q) \nonumber\\
	T_5 & = & 2 b_2 b_3 [ D(q,p) - x D(p_{K^*}, p)] \nonumber\\
	T_6 & = & 2 b_2 b_4 [ -D(q,q) + x D(q,p_{K^*})] \nonumber \\
	T_7 & = & -2 F b_3 b_4 D(p,q)
\eer
while
\ber
	H^{\mu\nu} {A^2}_{\mu\nu} &=& 
4c_1c_2[p_1 \cdot p_B p_2 \cdot p_{K^*}-p_2 \cdot p_B p_1 \cdot p_{K^*}]\\
\eer
and
\be
	g_{\mu\nu} H^{\mu\nu} = \sum S_i
\ee
\ber
	S_1 & = & -2 b_1^2 [m_{K^*}^2 m_B^2 - (p_B \cdot p_{K^*})^2] \nonumber\\
	S_2 & = & -3 b_2^2 \nonumber\\
	S_3 & = & F b_3^2 p^2\nonumber \\
	S_4 & = & b_4^2 q^2 \nonumber\\
	S_5 & = & 2 b_2 b_3 [q \cdot p - x p_{K^*} \cdot p] \nonumber\\
	S_6 & = & 2 b_2 b_4 [ -q^2 + x p \cdot q]\nonumber \\
	S_7 & = & -2 F b_3 b_4 p \cdot q
\eer
where
\ber
	F & = & \left[-q^2 + \frac{(p_{K^*} \cdot q)^2}{{m_{K^*}}^2} \right]
\nonumber \\
	x & = & \frac{p_{K^*} \cdot q}{{m_{K^{*}}^2 }}
\eer

Finally for {$K^*$ decay} the interference term is, with $p_1=p_B$,
$p_2=p_{K^*}$ and $q=p_1-p_2$,

\ber
	x_1 & = & 2b_1 [m_B^2 p_u \cdot p_2-p_1 \cdot p_u p_2 \cdot p_1]\nonumber \\
	x_2 & = & - b_2 \left[ 2 p_1 \cdot p_u 
		+ \frac{D(p_2,p_2)}{M_2^2}\right] b \nonumber\\
	x_3 & = & -  b_3 \left[ D(p_1+p_2,q) - \frac{p_2 \cdot q}{M_2^2}
		  D(p_1+p_2,p_2)\right] \nonumber\\
	x_4 & = & - b_4 \left[ -D(q,q) + \frac{p_2 \cdot q}{M_2^2} 
		D(q,p_2)\right] \nonumber\\
	y_1 & = & -3b_2 \nonumber\\
	y_2 & = 
         & b_3 \left[(p_1+p_2)\cdot q - (p_1+p_2)\cdot
        p_2\frac{p_2 \cdot q}{M_2^2} \right] \nonumber\\
	y_3 & = &  b_4\left[ q^2 - q 
\cdot p_2\frac{p_2 \cdot q}{M_2^2}\right] \nonumber\\
	D(A,B) & = & A \cdot p_1 \, B \cdot p_u + A \cdot p_u \, B 
\cdot p_B - p_u \cdot p_1 \, A \cdot B.  	
\eer

\end{document}